\documentclass{aa}

\usepackage{graphicx}
\usepackage{xcolor}
\usepackage{txfonts}
\usepackage[colorlinks=true,
            linkcolor=blue,
            citecolor=blue,
            urlcolor=blue,
            breaklinks=True]{hyperref} 
\usepackage{amsmath}

\newcommand{\pipe}{\mbox{\texttt{rPICARD}}}
\newcommand{\repo}{\url{https://bitbucket.org/M_Janssen/Picard}}
\usepackage{array}
\newcolumntype{P}[1]{>{\centering\arraybackslash}p{#1}}

\begin{document}

   \title{\pipe{}: A CASA-based calibration pipeline for VLBI data}

   \subtitle{Calibration and imaging of 7\,mm VLBA observations of the AGN jet in M87}

   \author{M.~Janssen
          \inst{1}
          \and
          C.~Goddi
          \inst{1,2}
          \and
          I.~M.~van~Bemmel
          \inst{3}
          \and
          M.~Kettenis
          \inst{3}
          \and
           D.~Small
          \inst{3}
          \and
          E.~Liuzzo
          \inst{4}
          \and
          K.~Rygl
          \inst{4}
          \and
          I.~Mart\'i-Vidal
          \inst{5}
           \and
          L.~Blackburn
          \inst{6}
          \and
          M.~Wielgus
          \inst{6,7}
          \and
          H.~Falcke
          \inst{1}
          }

   \institute{Department of Astrophysics/IMAPP, Radboud University Nijmegen, PO Box 9010, 6500 GL Nijmegen, The Netherlands\\
              \email{M.Janssen@astro.ru.nl}
         \and
             ALLEGRO/Leiden Observatory, Leiden University, PO Box 9513, NL-2300 RA Leiden, The Netherlands
         \and
             Joint Institute for VLBI ERIC, Dwingeloo, Postbus 2, 7990 AA, The Netherlands
         \and
             Istituto di Radioastronomia (INAF - IRA), Via P. Gobetti 101, I-40129 Bologna, Italy
         \and
             Observatorio de Yebes - IGN, Cerro de la Palera S/N, 19141 Yebes (Guadalajara), Spain
         \and
             Harvard-Smithsonian Center for Astrophysics, 60 Garden St., Cambridge, MA 02138, USA
        \and
            Black Hole Initiative at Harvard University, 20 Garden St., Cambridge, MA 02138, USA
             }

   \date{Received tbd; accepted tbd}

 
  \abstract 
   {The Common Astronomy Software Application (CASA) software
   suite, which is a state-of-the-art package for radio astronomy, can now reduce
   very long baseline interferometry (VLBI) data with the recent addition of a fringe fitter.}
   {Here, we present the Radboud PIpeline for the Calibration of high Angular Resolution Data (\pipe{}), which is
    an open-source VLBI calibration and imaging
   pipeline built on top of the CASA framework. The pipeline is capable of reducing data from different
   VLBI arrays. It can be run non-interactively after only a few non-default input parameters are set and
   delivers high-quality calibrated data.
   CPU scalability based on a message-passing interface (MPI) implementation
   ensures that large bandwidth data from future arrays can be processed within reasonable
   computing times.}
   {Phase calibration is done with a Schwab-Cotton fringe fit algorithm.
   For the calibration of residual atmospheric effects,
   optimal solution intervals are determined based on the signal-to-noise ratio (S/N) of the data for each scan.
   Different solution intervals can be set for different antennas in the same scan
   to increase the number of detections in the low S/N regime.
   These novel techniques allow \pipe{} to calibrate data from different arrays, including high-frequency and low-sensitivity arrays.
   The amplitude calibration is based on standard telescope metadata, and a robust algorithm can solve for atmospheric opacity
   attenuation in the high-frequency regime. Standard CASA tasks are used for CLEAN imaging and self-calibration. 
   }
   {In this work we demonstrate the capabilities of \pipe{} by calibrating and imaging
   7\,mm Very Long Baseline Array (VLBA) data of the central radio source in the M87 galaxy.
   The reconstructed jet image reveals a complex collimation profile and edge-brightened structure,
   in accordance with previous results. A potential counter-jet is detected that has 10\,\% of the
brightness of the   approaching jet. This constrains jet speeds close to the radio core to about
   half the speed of light for small inclination angles.}
   {}


   \maketitle
%

\section{Introduction}
\label{intro}

\nocite{eht-paperI}
\nocite{eht-paperII}
\nocite{eht-paperIII}
\nocite{eht-paperIV}
\nocite{eht-paperV}
\nocite{eht-paperVI}

Very long baseline interferometry (VLBI) is an astronomical observing technique used to study radio sources at very high angular resolution, down to milli- and
micro-arcsecond scales.
Global VLBI uses a network of
radio telescopes as an interferometer to form
a virtually Earth-sized telescope. 
The large distances between telescope sites impose the need of recording the radio wave signals and the use of independent and very precise 
atomic clocks at individual
VLBI stations, 
which allows a cross-correlation of the signals between all pairs of antennas
post facto.
The lack of real-time synchronization, the
typical sparsity of VLBI arrays, and the fact that the signals
received by each ground station are distorted by unique local atmospheric
conditions make the process of VLBI data calibration
especially challenging. At the correlation stage, these effects
are partially corrected for with a model of station locations, source positions,
the Earth orientation and atmosphere, tides, ocean loading, and relativistic
corrections for signal propagation \citep[see, e.g.,][]{Sovers1998,whitebook,bluebook}. 
These models are never perfect, however, and
the data must be calibrated in a post-correlation stage to correct for residual errors.
All further references to calibration procedures in this manuscript implicitly refer to post-correlation calibration.

While the Astronomical Image Processing System
(AIPS) \citep[e.g.,][]{Greisen2003} has been the standard package to calibrate
radio-interferometric and VLBI datasets, its successor, the Common Astronomy
Software Application (CASA) \citep{McMullin2007} package
has become the main tool for the calibration and analysis of connected element radio-interferometric data in recent years.
Nevertheless, AIPS continued to be the standard package for VLBI data reduction because CASA was missing 
a few key VLBI calibration functions, most notably, a fringe fitting task.
The missing functionalities have now been added to CASA
and the calibration framework has been augmented
by a global fringe fitting task through an initiative from the \href{https://blackholecam.org/}{BlackHoleCam} (BHC) \mbox{\citep{Goddi2017}} project
in collaboration with the Joint Institute for VLBI ERIC (JIVE) \citep{Bemmel2018}.
This task is based on the \mbox{\citet{Schwab1983}} algorithm and is similar to the FRING task in AIPS.
CASA presents some clear advantages over AIPS: 
a) an intuitive IPython interface implies a low learning curve for the new ‘python generation' of radio astronomers \citep{Momcheva2015};
b) software and data structure are designed to facilitate batch processing, providing much more control and flexibility
for pipeline-based data processing compared to AIPS, even when combined with the ParselTongue python framework \citep{Kettenis2006}, 
and c) strong community support,
largely by the Atacama Large Millimeter/submillimeter Array (ALMA) \citep{Wootten2009}
plus Karl G. Jansky Very Large Array (VLA) \citep{Thompson1980} userbases,
ensures the development and maintenance of a healthy software,
with quick bug detection and adjustments to the most recent needs of the community.
Under these conditions, it is natural to expect that CASA will soon become the standard package for VLBI data processing as well.

While traditionally only raw data taken by a radio interferometer were delivered to the principle investigator (PI), 
new-generation facilities such as ALMA provide raw data along with calibration tables obtained by running automated calibration pipelines. 
Next-generation facilities such as the Square
Kilometre Array (SKA) \citep[e.g.,][]{Dewdney2009} will generate much larger raw data volumes and only 
fully pipeline-calibrated datasets and images will be delivered to the PIs. 
For VLBI experiments, the number of participating stations is typically much
smaller than for connected interferometers. This leads to much smaller datasets, so that it is feasible to pass all the data on to the PI for post-correlation
calibration. A notable exception is the European VLBI Network (EVN) \citep[e.g.,][]{Porcas2010},
which provides users with a set of pipeline-generated calibration files and diagnostics \citep{Reynolds2002}.
The main purpose of this pipeline is a quick assessment of the quality and characteristics
of a dataset.
Advances in data-recording rates and wide-field VLBI capabilities will make it increasingly difficult
to reduce VLBI data interactively on single personal machines in the near future.
This necessitates data-handling methods, where computing power scales with available hardware.
Data reduction pipelines are an attractive solution to this problem, as they promote
reproducibility of scientific results and circumvent difficulties of VLBI data reduction.
As a byproduct, this will also attract more astronomers to the field of VLBI.

We have developed a highly modular, message-passing interface (MPI)-parallelized, fully automated VLBI calibration pipeline based on CASA,
called the Radboud PIpeline for the Calibration of high Angular Resolution Data (\pipe{}).\footnote{
The open-source \pipe{} software is hosted on \repo{}.
The pipeline is full dockerized (\url{https://www.docker.com}).}
The purpose of the pipeline is to provide science-ready data and thereby make VLBI more accessible to non-experts in the community.
It should be noted that VLBI data is prone to a large variety of data corruption effects.
Some of these effects cannot be remedied with calibration techniques and may escape the flagging
(removal of corrupted data) methods of \pipe{}.
Depending on the severity of these effects and the required quality of the data for scientific analysis, 
user interaction to address the data issues may be inevitable.
For these cases,  the verbose diagnostics, tuneability, and interactive capabilities of \pipe{} can be used to
obtain the required data quality.
\pipe{} v1.0.0 is able to handle data from any VLBI array when the raw data are in FITS-IDI\footnote{
See \url{https://fits.gsfc.nasa.gov/registry/fitsidi/AIPSMEM114.PDF}
for a description of the FITS-IDI data format.}
or MeasurementSet (MS)\footnote{See
\url{https://casa.nrao.edu/Memos/229.html}
for a description of the current MeasurementSet format.} format.

So far, EVN, Very Long Baseline Array (VLBA) \mbox{\citep{Napier1994}}, 
Global Millimeter VLBI Array (GMVA) \citep{Krichbaum2006}, and Event Horizon Telescope 
(EHT) \citep{eht-paperII} datasets have successfully been calibrated and imaged.
Phase-referencing and simple polarization calibration are supported. Future releases will be able to also handle spectral line observations.
CASA-based pipelines are already used for the reduction of ALMA and VLA data\footnote{See 
\url{https://casa.nrao.edu/casadocs-devel/stable/pipeline}.}
and with \pipe{}, a CASA-based VLBI pipeline is now available as well.

We describe the general CASA calibration scheme in Section~\ref{casacalib} and the structure of the \pipe{} framework in
Section~\ref{picode}. A description of the pipeline calibration strategies follows in Section~\ref{picalib}. The implementation
of CASA-based automated imaging and self-calibration routines in \pipe{} are specified in Section~\ref{imagscal}.
A verification of the pipeline based on VLBA data is presented in Section~\ref{verification}. Section~\ref{futurefeatures}
gives an overview of future features, and a concluding summary is given in Section~\ref{summary}.

\section{CASA calibration framework}
\label{casacalib}
CASA makes use of Jones matrices \citep{Jones1941} and the Hamaker-Bregman-Sault measurement equation \citep{Hamaker1996}
to calibrate full Stokes raw complex VLBI visibilities formed at the correlator.
In this framework, visibility measurements $\vec{\mathcal{V}}_{mn}(t, \nu)$ at a time $t$ and frequency $\nu$,
on a baseline $m$-$n$, can be represented in vector notation as
\begin{equation}
\vec{\mathcal{V}}_{mn} (t, \nu) = 
\left\langle \,
 \begin{pmatrix}
   V_{ma} V_{na}^*   \\
   V_{ma} V_{nb}^*   \\
   V_{mb} V_{na}^*   \\
   V_{mb} V_{nb}^*  
 \end{pmatrix} \, \right\rangle \; .
\label{hms-vector}
\end{equation}
Here, $V_{xy}$ represents the measured complex voltages at station $x$ along signal path $y$,
the star denotes a complex conjugate, and the angle brackets indicate an integration over small
time and frequency bins at the correlator. For circular telescope feeds, for example, right circular polarization
(RCP) and left circular polarization (LCP) signals are measured by the $a$ and $b$ signal paths, respectively.
Therefore the four rows of $\vec{\mathcal{V}}_{mn}(t, \nu)$ would represent the RR, RL, LR, and LL correlations.
The voltages $V$ were recorded at a specific time $t$ and frequency $\nu$. Their dependence 
on these quantities is not explicitly shown here for the sake of a simpler notation.
For the remainder of this work, the explicit time and frequency dependence of all visibility-related quantities are omitted.

Ideal, uncorrupted visibilities
$\vec{\mathcal{V}}_{mn}^\mathrm{true}$,
which would be obtained from a perfect measurement device,
are related to the measured visibilities
$\vec{\mathcal{V}}_{mn}^\mathrm{obs}$ through a $4\times4$ matrix $\mathcal{J}_{mn}$, which contains all accumulated
measurement corruptions on baseline $m$-$n$:
\begin{equation}
\label{mseq1}
\vec{\mathcal{V}}_{mn}^\mathrm{obs} = \mathcal{J}_{mn} \vec{\mathcal{V}}_{mn}^\mathrm{true} \;.
\end{equation}
Equation~\ref{mseq1} assumes that telescopes are linear measurement devices, therefore no higher-order
terms of $\mathcal{J}_{mn}$ are considered.
Examples of corruptions that factor into $\mathcal{J}_{mn}$ are antenna gain errors, antenna bandpasses, and
atmospheric phase distortions. We can denote the individual constituents of $\mathcal{J}_{mn}$ by $J_{mn}^k$,
where each index $k$ represents a different corruption effect. The order of $k$ should be equal
to the reverse order in which the corruptions occur along the signal path, that is, first the 
instrumental effects introduced by the signal recording, then effects from the receiving
elements, and finally atmospheric signal corruptions.
The combined effect of all corruption effects
can be represented as
\begin{equation}
\mathcal{J}_{mn} = \prod_k J_{mn}^k \;.
\end{equation}
Only antenna-based corruption effects are typically removed in the calibration process, as
baseline-based effects have a much smaller magnitude and are more difficult to determine.
This means that $J_{mn}^k$ can be rewritten as $J_{m}^k \otimes \left(J_{n}^k\right)^*$,
where the $\otimes$ operator represents the tensor product.
These $J_{m}^k$ factors correspond to $2\times2$ Jones matrices.
Now, Equation~\ref{mseq1} can be written as 
\begin{equation}
\label{mseq2}
\vec{\mathcal{V}}_{mn}^\mathrm{obs} = \prod_k \left[ J_{m}^k \otimes \left(J_{n}^k\right)^* \right] \vec{\mathcal{V}}_{mn}^\mathrm{true} \;.
\end{equation}

CASA keeps the complex visibility data, together with auxiliary metadata (antennas, frequency setup,
system temperatures, etc.),
stored locally in a contained form: as binary tables that make up an MS.
The calibration philosophy is to perform incremental calibration based on the inverse of Equation~\ref{mseq2} with separate
tables for each corruption effect, containing the calibration solutions.
These calibration tables have the same structure
as the MS, are also stored as self-describing binary data, and are applied in Jones matrix form.
When solving for a new calibration, previous calibration solutions for other corruption effects can be 
applied `on-the-fly', meaning that the data are calibrated while they are passed to the solver.
The new calibration solutions are therefore relative to the previous ones. Typically, the dominant 
data corruption effects are calibrated out first. This can be an iterative 
process if different corruption effects are not sufficiently independent.
If no good calibration solution can be obtained, a flagged solution will be written in
the calibration table. This generally happens when the S/N of the data is not sufficient to obtain
a reliable calibration. Applying flagged solutions to the MS will cause the corresponding data to be flagged.
The final goal of the calibration process is to obtain a close representation of $\vec{\mathcal{V}}_{mn}^\mathrm{true}$
by applying all calibration tables to the measured $\vec{\mathcal{V}}_{mn}^\mathrm{obs}$.

It should be noted that station-based calibration is generally an overdetermined problem: For $N$ antennas,
there are $N \left( N - 1 \right)/2$ baseline-based visibility measurements.

CASA keeps track of \textit{SIGMA} and \textit{WEIGHT} columns corresponding to the visibility data.
\textit{SIGMA} represents the noise within a frequency channel of width $\Delta \nu$
in a time bin $\Delta t$. For a single complex cross-correlation visibility data point, this noise is
given by
\begin{equation}
\label{sigma}
\sigma = \frac{1}{\sqrt{ 2 \Delta \nu \Delta t}} \;.
\end{equation}
CASA will take into account correlator weights when estimating $\sigma$.
For example, for data from the DiFX correlator \citep{Deller2007}, weights are determined based on the amount of valid data present
in each integration bin $\Delta t$ when initializing the \textit{SIGMA} column.
After this, \textit{SIGMA} will only be modified when data are averaged according to the changes
in $\Delta \nu$ and $\Delta t$. 
The \textit{WEIGHT} column is used to weight data according to their quality when averaging within
certain bins.
The column is initialized based on the initial \textit{SIGMA} column as $\sigma^{-2}$ for each visibility.
After initialization, the product of all applied station-based gains
will modify the weights of the baseline-based visibilities. 
For instance, high system temperature values
and channels that roll off at the edge of the bandpass will be down-weighted in this process.
For frequency-dependent weights, the CASA \textit{SIGMA\_SPECTRUM} and \textit{WEIGHT\_SPECTRUM}
columns are used.

The spectral setup of the MS data format consists of 
spectral windows (spws) that are subdivided into frequency channels.
Channels are formed at the correlator and determine the
frequency resolution of the data.
Spectral windows correspond to distinct
frequency bands, equivalent to AIPS intermediate frequencies (IFs).
These frequency bands are usually formed by the heterodyne receivers of the telescopes when
the high-frequency sky signal is mixed with a local oscillator signal.
The resulting frequency down-conversion enables analog signal processing
\citep{Thompson1980}.
Data from different spectral windows
have usually passed through different electronics, therefore instrumental effects
must often be calibrated for each spw separately before 
the data from the full observing bandwidth are combined.

\section{Code structure of \pipe{}}
\label{picode}
\pipe{} is a software package for the calibration and imaging of VLBI data based on
CASA. This section describes the most important features of the pipeline source code,
in particular, code philosophy (§\,\ref{code:philo}), input and output (§\,\ref{code:IO}),
handling of metadata (§\,\ref{subsec_meta}), interactive capabilities (§\,\ref{code:interactive}),
handling of data flag versions (§\,\ref{code:flag}), and the
MPI implementation (§\,\ref{code:mpi}).

\subsection{Code philosophy}
\label{code:philo}
The source code is written based on a few maxims:
\begin{enumerate}
\item Every parameter can be set in input files; nothing is hard coded. \vspace{0.05cm}
\item No parameter needs to be set by hand because of sensible or self-tuning default values. \vspace{0.05cm}
\item By default, every run of the pipeline is tracked closely with
      very verbose diagnostic output. \vspace{0.05cm}
\item Every step of the pipeline can always be repeated effortlessly.
\end{enumerate}
This allows experienced users to tune the pipeline to their needs, while new users
will be able to use \pipe{} to learn about the intricacies of VLBI data reduction.
Similarly, the pipeline can be used either for a quick-and-dirty analysis or
to obtain high-quality calibrated data ready for scientific analysis.

\subsection{Input and output}
\label{code:IO}
Input parameters are read in from simple configuration text files.
The raw input visibility data from the correlator can either be an MS or a set of FITS-IDI files,
which will be imported as MS into CASA with the \textit{importfitsidi} task.
Optional metadata files are read in when available as described in Section~\ref{subsec_meta}.

Additional command-line arguments can be used to enable the interactive mode, and to
select which pipeline steps are to be repeated when the user experiments
with different non-default calibration parameters (e.g., S/N cutoffs,
interpolation methods of solutions, and selection of calibrator sources).

Finally, a calibrated MS with user-defined spectral and time averaging is produced.
For backward comparability with older radio astronomy software packages, UVFITS files
will be created from the calibrated MS as well.\footnote{See
\url{ftp://ftp.aoc.nrao.edu/pub/software/aips/TEXT/PUBL/AIPSMEM117.PS}
for a description of the UVFITS file data format.}
\footnote{
It should be noted that the UVFITS files produced by CASA can currently not be read into AIPS.
This was tested with CASA version 5.4 and AIPS version 31DEC18. The issue is related to different
formatting conventions of the `AIPS AN' extension of the UVFITS file.
Difmap \citep{Shepherd1994} is able to read in the CASA UVFITS files without problems.}

\subsection{Metadata}
\label{subsec_meta}
A priori knowledge is crucial for the calibration of VLBI data.
\pipe{} looks for the following files and reads them in automatically as metadata if they exist:
\begin{itemize}
\item Standard ANTAB\footnote{The ANTAB format is the current standard used for VLBI flux calibration.
See \url{http://www.aips.nrao.edu/cgi-bin/ZXHLP2.PL?ANTAB}
for more information.} files containing system temperatures ($T_\mathrm{sys}$) and station gains,
which are used for the amplitude calibration (Section~\ref{apcal_section}). Some correlators
will already attach a $T_\mathrm{sys}$ table to the raw visibility data themselves.
Otherwise, custom scripts are used to attach the ANTAB data to the visibilities.
\item Files containing the receiver temperature information of the stations, which increases
the accuracy of the correction for the signal attenuation caused by
the opacity of the Earth's troposphere. This additional opacity correction is recommended for
high-frequency observations that do not measure opacity-corrected $T_\mathrm{sys}$ values (Section~\ref{apcal_section}).
This is the case for the switched-noise calibration method of the VLBA, for example.
\item Weather tables from the antenna weather stations, which are only needed if an additional
opacity correction is to be performed. Usually, these tables are already attached to the 
visibility data. If the weather information is only present in ASCII format, it will be 
attached to the visibilities with custom scripts.
\item Files containing flagging instructions either in the native CASA format or the
AIPS-compliant UVFLG format. Files with flagging instructions are typically compiled
from observer logs of the stations and are handed to PIs together with the correlated data.
\item Files containing models of the observed sources. When available, these source models can 
improve some calibration steps (e.g., fringe fitting).
Usually, no a priori source models are needed (Section~\ref{picalib}). 
\item A file with the mount-type corrections of the stations when they are not correctly
specified in the MS or FITS-IDI files. The mount-types of a station are important for the
polarization calibration, specifically, the feed-rotation angle correction (Section~\ref{phasecal}).
The feed rotation describes the rotation of the orthogonal polarization receiving elements of a telescope
as seen from the sky.
\end{itemize}

The naming conventions for each of the files are described in the
online documentation.\footnote{\url{https://bitbucket.org/M_Janssen/picard/src/master/documentation}.}

\subsection{Interactivity capabilities}
\label{code:interactive}
For a typical VLBI dataset, the basic flagging mechanisms (Section~\ref{flagging})
and fringe non-detections (Appendix~\ref{ffbasics}) should take care of bad data. 
For severely corrupted datasets, for example, with phase instabilities on very short timescales, 
data dropouts, or unrecoverable correlation errors, 
user-interaction is inevitable if a high-quality calibration is to be achieved.
In \pipe{}, an interactive mode is enabled with a simple command-line flag.
In this mode, the software waits for user input to advance to the next step.
This allows for a careful inspection and refinement through manual flagging and/or 
smoothing of calibration tables. Moreover, it is always possible to flag
poor visibilities that are responsible for erroneous calibration solutions and to quickly
repeat the calibration step. The pipeline provides an external module containing
functions for convenient post-processing of calibration tables. It
can be loaded into an interactive CASA session and be used in between 
interactive steps.

\subsection{Flag versions}
\label{code:flag}
\pipe{} keeps track of different flag versions of the visibility data in the MS
using the default lightweight CASA \texttt{.flagversions} extensions of the MS.
Three flagging journals are stored:
\begin{enumerate}
\item An initial blank flag version created immediately after the data have been loaded
for the first time.
\item A version with all a priori flags applied (Section~\ref{flagging}).
\item A final version after all calibration tables have been applied.
This version will include the a priori flags and flags based on failed
calibration solutions.
\end{enumerate}

These different flag versions are relevant when parts of the pipeline
are rerun for an optimization of calibration steps. The first version can be used
when starting from scratch with a new strategy, for instance, and the second version
can be used when only certain steps with a different set of
parameters are rerun.
The flag version can be specified at the start of a pipeline run.

   \begin{figure*}[h!]
   \centering
   \includegraphics[width=0.85\textwidth]{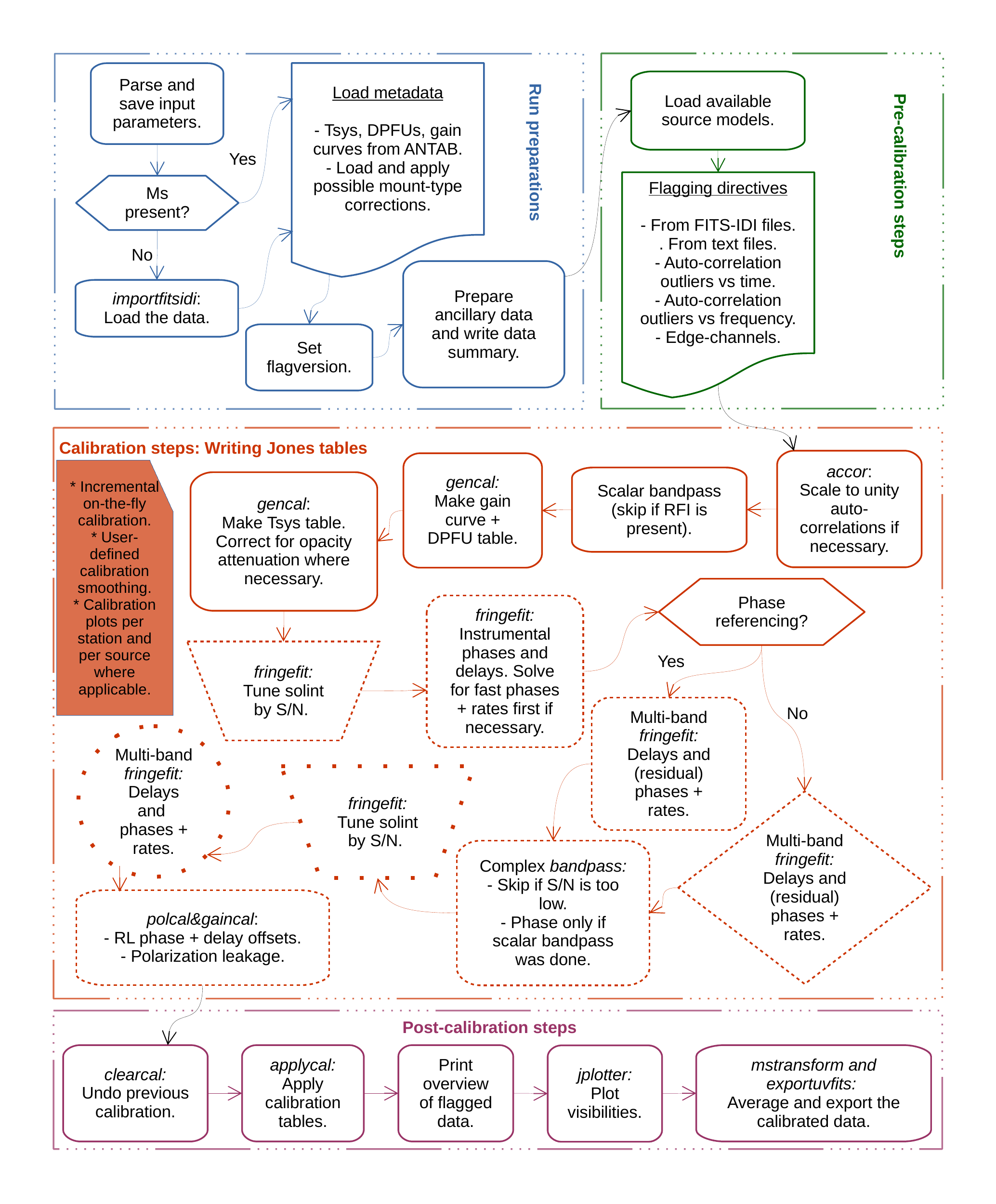}
   \caption{Overview of the \pipe{} calibration scheme.
            CASA tasks used by the pipeline are written in italics.
            In the orange block in the middle, distinctions are made based on 
            which sources are used to solve for the specific calibration tables
            and to which sources these tables are applied.
            Boxes with a solid border belong to calibration steps where
            all sources were used to obtain solutions. Dashed borders mean
            that only calibrators were used, and dotted borders correspond
            to solutions obtained from the science targets only. The line style of the borders only describes the source selection;
            an averaging of solutions from different sources is not implied.
            Rectangular boxes indicate solutions that are applied to all
            sources. Diamonds represent solutions applied to calibrators only,
            circles are used when the solutions are applied to science targets only,
            and trapezoids indicate intermediate calibration solutions that
            are not applied to the data.
            }
              \label{calib_flowchart}
    \end{figure*}

\subsection{MPI parallelization}
\label{code:mpi}
With the steadily increasing data-recording rates of VLBI arrays, it is necessary for
downstream calibration and analysis software packages to scale up with the available computing power, and in particular, to fully exploit the parallelism of modern multi-node, multi-core machines.
Within CASA, central processing unit (CPU) scalability is implemented through an MPI infrastructure.
The MS is partitioned into several sub-MSs that are virtually concatenated.

In \pipe{}, the MS is subdivided across scans. In this way, 
multiple workers can calibrate multiple scans simultaneously.
The most significant acceleration is achieved for the fringe fitting steps 
(Section~\ref{phasecal}), where the least-squares globalization steps to go from
baseline-based fringes to station-based solutions require significant
CPU time, especially for large bandwidths. For other CASA tasks, which
are internally parallelized, shorter computing times are achieved as well.
The CASA MPI implementation is still being developed, which means that more and more
CASA tasks will be upgraded with an MPI functionality in the future.
In Appendix~\ref{mpibench}, we present a test case that demonstrates
the significance of the CPU scalability.

An optional input parameter can be set to control the memory usage of 
the MPI servers: jobs will only be dispatched when enough memory is available.
This option is disabled by default because the memory monitoring and resource scheduling
will slightly decrease the performance of the pipeline. Moreover, there will be no
memory limitations for most VLBI datasets\footnote{Wide-field and very wide-bandwidth 
VLBI observations may be exceptions.}
on systems with a reasonable ratio of CPU cores to random-access memory (RAM).

\section{\pipe{} VLBI calibration procedures}
\label{picalib}

This section describes the calibration procedures employed by \pipe{}.
In accordance with standard VLBI data reduction,
the user has to specify suitable calibrator sources for the different calibration steps
in the pipeline input files.
For the calibration of
the phase bandpass and other instrumental phase corruption effects, bright, compact, 
flat-spectrum sources should be used. For phase-referencing,
a calibrator close to the science target on the sky is required. A linearly
polarized source tracked over a wide range of feed rotation angles is needed for
proper polarization calibration.

An overview of the overall calibration scheme is given in Fig.~\ref{calib_flowchart}.
The standard VLBI calibration techniques introduced in this section,
using ANTAB data and fringe fitting techniques, are referred to as model-agnostic
calibration to distinguish it from the model-dependent additional self-calibration introduced in Section~\ref{imagscal}.
This section presents flagging methods that are used by the pipeline to
remove poor visibilities in §\,\ref{flagging}.
§\,\ref{digicorr} describes the calibration for digitization effects,
§\,\ref{apcal_section} outlines the amplitude calibration,
§\,\ref{scalar-bpass} describes the amplitude bandpass correction,
§\,\ref{phasecal} presents the phase calibration framework,
and §\,\ref{polcalsect} outlines the methods used for 
the polarization calibration.
Amplitude calibration steps are done first because they will adjust the data weights and 
therefore guide the phase calibration solutions (Appendix~\ref{incrcalib_and_wt}).

\subsection{Flagging methods}
\label{flagging}
Poor data can severely inhibit the science return of VLBI data.
Some corruptions cannot be calibrated, and the affected data should be removed (flagged). Typically, these are data with a low S/N for
which no calibration can be obtained by bootstrapping from neighboring calibration solutions. Examples are data recorder failures
and edge channels without signal.
\pipe{} has several independent steps to deal with data flagging:
\begin{enumerate}
\item Flags from the correlator are applied, which are typically
complied from station log files, which describe when telescopes are off-source.
For the pipeline, correlator flags can come attached to the FITS-IDI input files, or as separate files
either as AIPS-compliant UVFLG flags or as a file with flag commands in the CASA format. UVFLG flags are first converted
into the CASA flag format with custom scripts.
\item Edge-channels can be flagged if the spectral windows
in the data
are affected by a roll-off at the edges.
The user can specify the number of channels to be flagged. The default is to flag the first and last 5\% of channels.
\item Users can add their own flagging statements
based on data examinations as flagging files for easy bookkeeping.
\item Finally, experimental flagging algorithms have been implemented in \pipe{}. These try to find
poor data based on outliers in autocorrelation spectra as a function of time and frequency
with respect to median amplitudes.

In the frequency space, an autocorrelation spectrum is formed by
averaging in time over scan durations. Outliers are identified if the running difference in autocorrelation amplitude across channels is 
larger than the median difference across all channels by a user-defined fraction. This analysis is done on a 
per scan, per antenna, per spectral window, and per parallel-hand correlation basis. 
The purpose is to identify poor data that are due to dead frequency channels
and radio-frequency interference (RFI).

Along the time axis, the autocorrelation data are averaged across all channels and all spectral windows. Then, the median
autocorrelation amplitude is found and outliers are identified if the amplitude of an autocorrelation
data point differs from the median by more than a user-defined fraction.
This analysis is made for each parallel-hand correlation and
each station separately. The purpose is to identify poor data that are due to recorder issues or 
antennas arriving late on source.

The derived flagging commands are compiled into flag tables and are applied to both the
auto- and cross-correlations.
These experimental flagging algorithms should be used with care as they have only been tested with 
(and successfully applied to) EHT data.
\end{enumerate}
The flagging instructions from each step are written as ASCII CASA flag tables
and are applied to the data with the CASA \textit{flagdata} task prior to each
calibration step.

\subsection{Sampler corrections}
\label{digicorr}

The analog signals measured by each station are digitized and recorded for later cross-correlation to form visibilities.
Erroneous sampler thresholds from the signal digitization stage are determined with the \textit{accor} CASA task.
Correction factors $g^\mathrm{accor}$ are derived based on how much the autocorrelation spectrum of each station deviates from unity.\footnote{For some datasets, this
correction is not necessary: The SFXC correlator already applies this correction for EVN data, for example.}

\subsection{Amplitude calibration}
\label{apcal_section}

The digital sampling of the received signals at each station causes a loss of information about the amplitude of the electromagnetic waves.
The lack of calibrator sources with a `known' brightness prevents a scaling of the amplitudes on all baselines to recover the
correct source flux densities; typical VLBI sources are resolved and variable, and accurate time-dependent a priori source models
are not available.
Instead, the system equivalent flux densities (SEFDs) of all antennas can be used to perform the amplitude calibration. The SEFD of an
antenna is defined as the total noise power, that is, a source with a flux density equal to the SEFD would have an S/N
of unity. It can be written as
\begin{equation}
\label{eq_sefd}
\mathrm{SEFD} = \frac{T_{\mathrm{sys}} e^{\tau}}{\mathrm{DPFU} \cdot \mathrm{gc}} \; .
\end{equation}
Here, $\mathrm{T_{sys}}$ is the system noise temperature in Kelvin and 
the $\mathrm{DPFU}$ is the degrees per flux unit factor in Kelvin per Jansky (Jy).\footnote{
The Jansky unit is defined as $10^{-26}$ watts per square meter per hertz.} The $\mathrm{DPFU}$
describes the telescope gain at an optimal elevation, relating a measured temperature to a flux.
The \mbox{$\mathrm{gc}=\mathrm{gc}(\mathrm{elevation})$} variable describes the telescope
gain-elevation curve, which is a function normalized to
unity that takes into account the changing gain of the telescope due to non-atmospheric (e.g., gravitational deformation) elevation effects. 
The $\tau$ factor is the atmospheric opacity, which attenuates the received signal.
A $e^{\tau}$ correction enters in the SEFD to represent the signal from above the atmosphere (before atmospheric attenuation).
In this way, source attenuation
will be accounted for in the amplitude calibration.

\begin{figure}[b]
   \centering
   \includegraphics[width=0.48\textwidth]{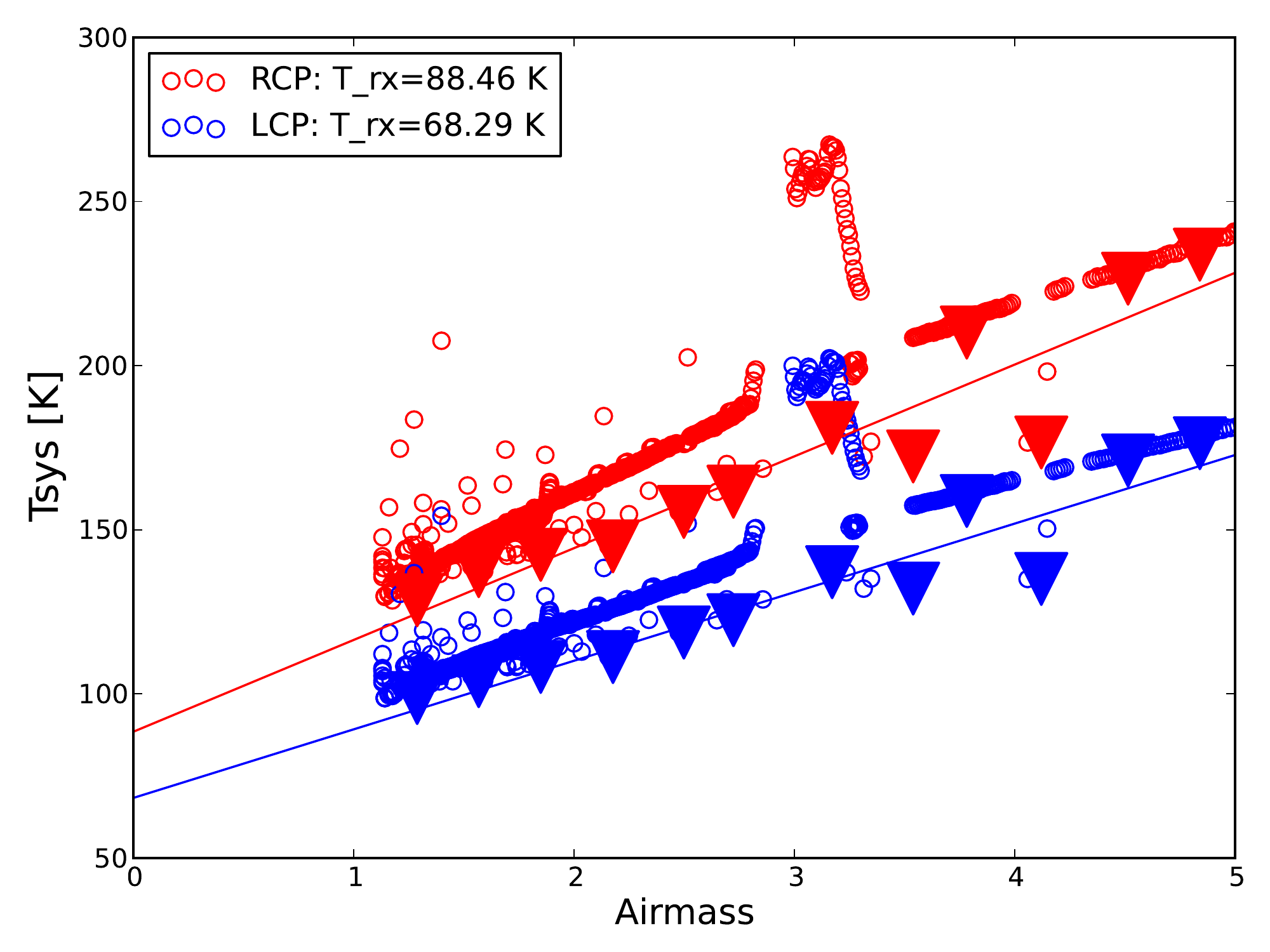}
      \caption{System temperature from the Brewster VLBA station at 7mm on 04 June 2013 as a function
      of airmass (given by $1/\sin{(\mathrm{elevation})}$) for the RCP and LCP receivers (red and blue open circles, respectively).
      Overplotted are solid triangles from the 
      lowest $T_\mathrm{sys}$ values in each bin of 0.3 airmass size, which are used to determine the receiver temperature 
      by fitting a straight line to the lower $T_\mathrm{sys}$ envelope (formed by the binning).
      Small $T_\mathrm{sys}$ variations induced by weather (e.g., the jump seen at an airmass of $\sim3$)
      do not affect the fit.
              }
         \label{tsysstar}
   \end{figure}

In the following, we describe how \pipe{} solves for the $e^{\tau}$ atmospheric opacity correction factor.
The system noise temperature can generally be written as 
\begin{equation}
\label{tsys}
T_\mathrm{sys} = T_\mathrm{rx} + T_\mathrm{sky} + T_\mathrm{spill} + T_\mathrm{bg} + T_\mathrm{loss} + T_\mathrm{source} \;,
\end{equation}
with $T_\mathrm{rx}$ the receiver temperature, $T_\mathrm{sky}$ the sky brightness temperature, $T_\mathrm{spill}$ the contribution from stray
radiation of the telescope surroundings, $T_\mathrm{bg}$ the cosmic microwave background and galactic background emissions, 
$T_\mathrm{loss}$ the noise contribution from losses in the signal path, and $T_\mathrm{source}$ the 
contribution from the source, which is usually smaller than the other noise temperatures.
Rewriting the sky brightness temperature in terms of the attenuated actual temperature of the atmosphere $T_\mathrm{atm}$ yields
\begin{equation}
\label{tsky}
T_\mathrm{sky} = \left( 1 - e^{-\tau} \right) T_\mathrm{atm} \; ,
\end{equation}
with $\tau = \tau_0 / \sin{(\mathrm{elevation})}$ the opacity at zenith $\tau_0$, corrected for the airmass at the elevation of the telescope, which is
given as $1/\sin{(\mathrm{elevation})}$. Using Equation~\ref{tsky}, the dominant terms of Equation~\ref{tsys} can be written as
\begin{equation}
\label{tsys2}
T_\mathrm{sys} \simeq T_\mathrm{rx} + \left( 1 - e^{-\tau} \right) T_\mathrm{atm} \; .
\end{equation}
The standard hot-load calibration method of high-frequency observatories will measure an opacity-corrected
system temperature $T_\mathrm{sys}^{*} \equiv T_\mathrm{sys} e^{\tau}$ directly \citep{Ulich1976},
while many other telescopes, such as the VLBA stations, will use a switched-noise method \citep[e.g.,][]{Oneil2002} that does not
correct for the atmospheric attenuation. For millimeter observations, where the signal attenuation is significant, \pipe{} will solve for
$e^{\tau}$ for each opacity-uncorrected $T_\mathrm{sys}$ measurement, using Equation~\ref{tsys2}.
The receiver temperatures can either be inserted beforehand from a priori knowledge about the station frontend, or the pipeline will
estimate $T_\mathrm{rx}$ based on the fact that $T_\mathrm{sys}$ should converge to $T_\mathrm{rx}$ toward zero airmass. This 
is done by fitting the lower envelope (to be robust against $\tau$ variations) of system temperature versus airmass, 
exemplified in Fig.~\ref{tsysstar}.
This method was adopted from \citet{martividal2012}.
The other unknown in Equation~\ref{tsys2}, the atmospheric temperature corresponding to a system temperature measurement, 
is estimated based on the \citet{Pardo2001} atmospheric model
implementation in CASA with input from local weather stations.

With these assumptions, \pipe{} can determine $T_\mathrm{sys}^{*}$ for the SEFD calculations from each $T_\mathrm{sys}$ measurement.
Outlier values are removed by smoothed interpolation to make the fitting robust even in poor weather conditions.
It is possible to perform the additional opacity correction only for a subset of stations in the VLBI array.

Within CASA, polynomial gain curves from the ANTAB files are first multiplied by the DPFUs, then the square root is taken, and
finally, the polynomial coefficients are refit with respect to 90 degrees elevation instead of zero. In this way, VLA-type
gain curves are formed. These gain curves and the system temperatures are converted into standard CASA amplitude calibration tables
with the \textit{gencal} task.

On a baseline between stations $m$ and $n$, the amplitude calibration and sampler correction $J^\mathrm{accor}$ will adjust
visibility amplitude $\vec{\mathcal{V}}_{mn}^\mathrm{amp}$ as 
\begin{equation}
\label{apcal}
\vec{\mathcal{V}}_{mn}^\mathrm{true,\,amp} = J^\mathrm{accor}_{m} \, J^\mathrm{accor}_{n} \sqrt{\mathrm{SEFD}_m \mathrm{SEFD}_n} \vec{\mathcal{V}}_{mn}^\mathrm{obs,\,amp} \; .
\end{equation}

\subsection{Scalar bandpass calibration}
\label{scalar-bpass}
The data from each spectral window pass through a bandpass filter.
These filters never have a perfectly rectangular passband, which \pipe{} corrects for
by determining an accurate amplitude (scalar) bandpass calibration solutions
from the autocorrelations within each spectral window. The autocorrelations are formed by correlating the
signal of a station with itself, that his, they are the Fourier transform of the power spectrum and do not contain phase information.
Because their S/N is higher, a scalar bandpass solution can be computed
for each scan by averaging the data over the whole scan duration and taking
the square root of normalized per-channel amplitudes.
This is done with a custom python script using basic CASA tools.
The user should disable this step when the autocorrelations are affected by RFI.
In this case, the complex bandpass calibration (Section~\ref{bpass}) can be used to correct the amplitudes.

\subsection{Phase calibration}
\label{phasecal}
The large distances between telescopes in VLBI arrays mean that independent local
oscillators are required at the stations. The large distances also cause the signals to be corrupted by independent atmospheres.
When visibilities are formed at correlators, sophisticated models are used 
to compute station and source positions and to align the wavefronts
from pairs of stations \citep[e.g.,][]{whitebook}.
However, the correlator models are never perfect, and residual phase corruptions will still be present
in the data. These are phase offsets, phase slopes with frequency (delays), and phase slopes with time (rates or fringe rates).
The principal task for any VLBI calibration software is to calibrate these errors 
out with a fringe fitting process \citep[e.g.,][]{bluebook}, so that the data can be averaged in
time and frequency to facilitate imaging and model-fitting in the downstream analysis. 

In practice, fringe fitting is used to correct for both instrumental and atmospheric effects.
Instrumental effects occur as data from different spectral windows pass through different signal paths,
causing phase and delay offsets between the spectral windows.
This can be modeled as a constant or slowly varying effect
over time, and can be solved for by fringe fitting each spectral window individually in single-band fringe fit steps.
The atmosphere causes time-varying phases, delays, and rates. Delays are typically stable over scans,
unless the signal path changes, for example, when clouds pass by.\footnote{
Gross fringe offsets are usually removed by a priori correlator models.
} Phases and rates, as they describe the change 
in phase with time, can vary on timescales given by the atmospheric coherence time.
These coherence times are very short, $\mathcal{O}(\mathrm{seconds})$, for millimeter observations.
In the centimeter (cm) regime, the coherence times are $\mathcal{O}(\mathrm{minutes})$.
Atmospheric effects are taken out over wide
bandwidths to maximize the S/N, that is, data from multiple spectral windows are used in multi-band fringe fit steps.

\pipe{} employs multiple fringe fitting steps using 
the recently added \textit{fringefit} CASA task, which is based on the \citet{Schwab1983} algorithm.
First, optimal solution intervals for atmospheric effects are determined for the 
calibrator sources (Section~\ref{solints}). For millimeter observations, these solution intervals are used
to perform a coherence calibration, where intra-scan phase and rate offsets are corrected
to increase the coherence times for the calibrator sources (Section~\ref{incrcoher}).
The next step is to correct for instrumental phase and delay offsets (Section~\ref{instrphdela}).
Then, residual atmospheric effects are corrected for the calibrator sources (Section~\ref{mbcal}),
which finalizes the fringe calibration for the calibrators.
If the S/N of the calibrator scans is good enough, corrections for the antenna phase bandpass
can be obtained (Section~\ref{bpass}).
With all instrumental effects taken out, the phases of the weaker science targets
are calibrated (Section~\ref{mbphsci}).
More details about fringe fitting in CASA are given in Appendix~\ref{ffbasics}.
It should be noted that all fringe fit solutions are obtained after the
geometric feed rotation angle phase evolution has been corrected (Appendix~\ref{mounttypes}).

\subsubsection{Finding the optimal solution intervals for the calibrator sources}
\label{solints}

   \begin{figure}
   \centering
   \includegraphics[width=0.5\textwidth]{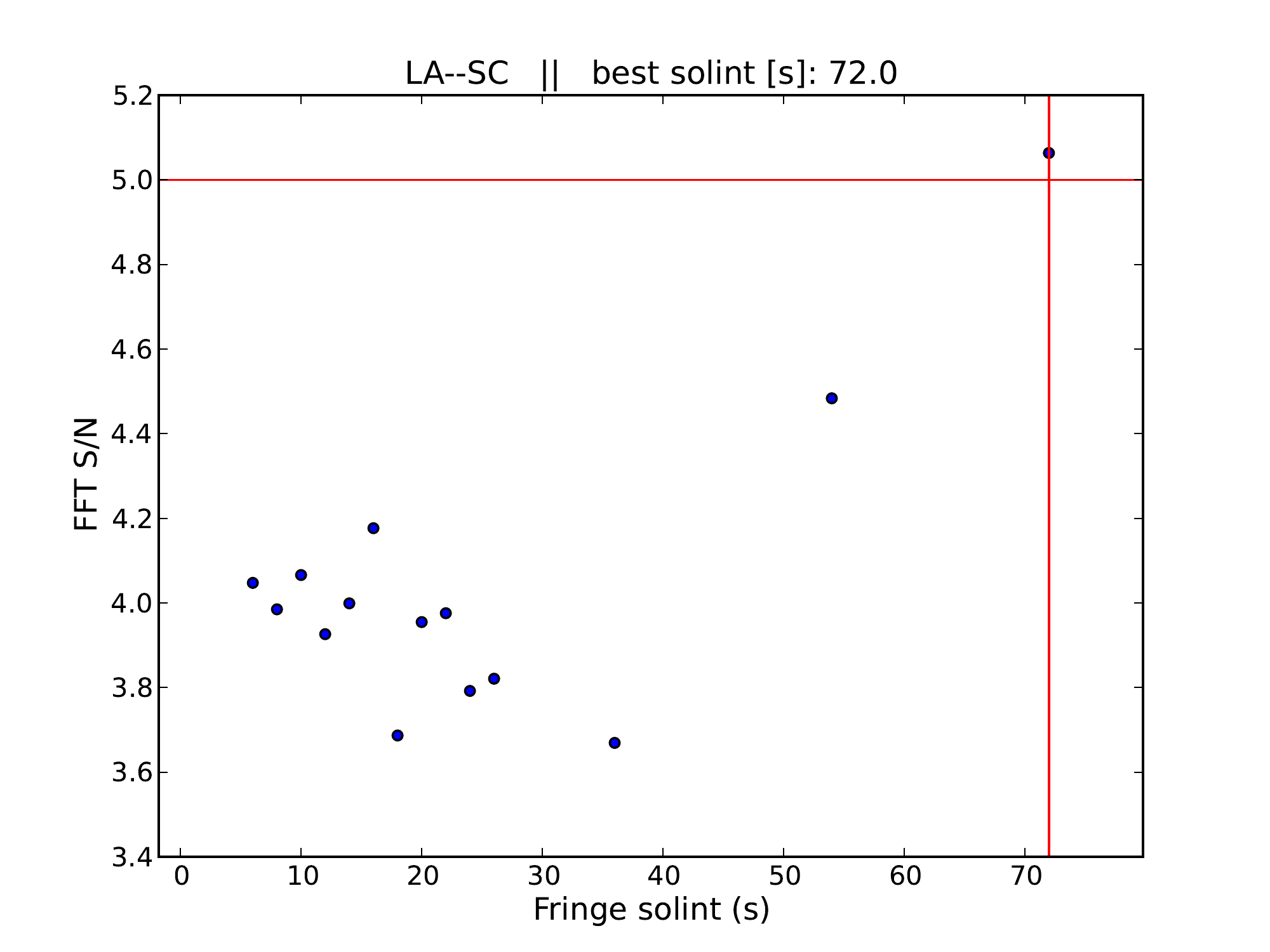}
   \includegraphics[width=0.5\textwidth]{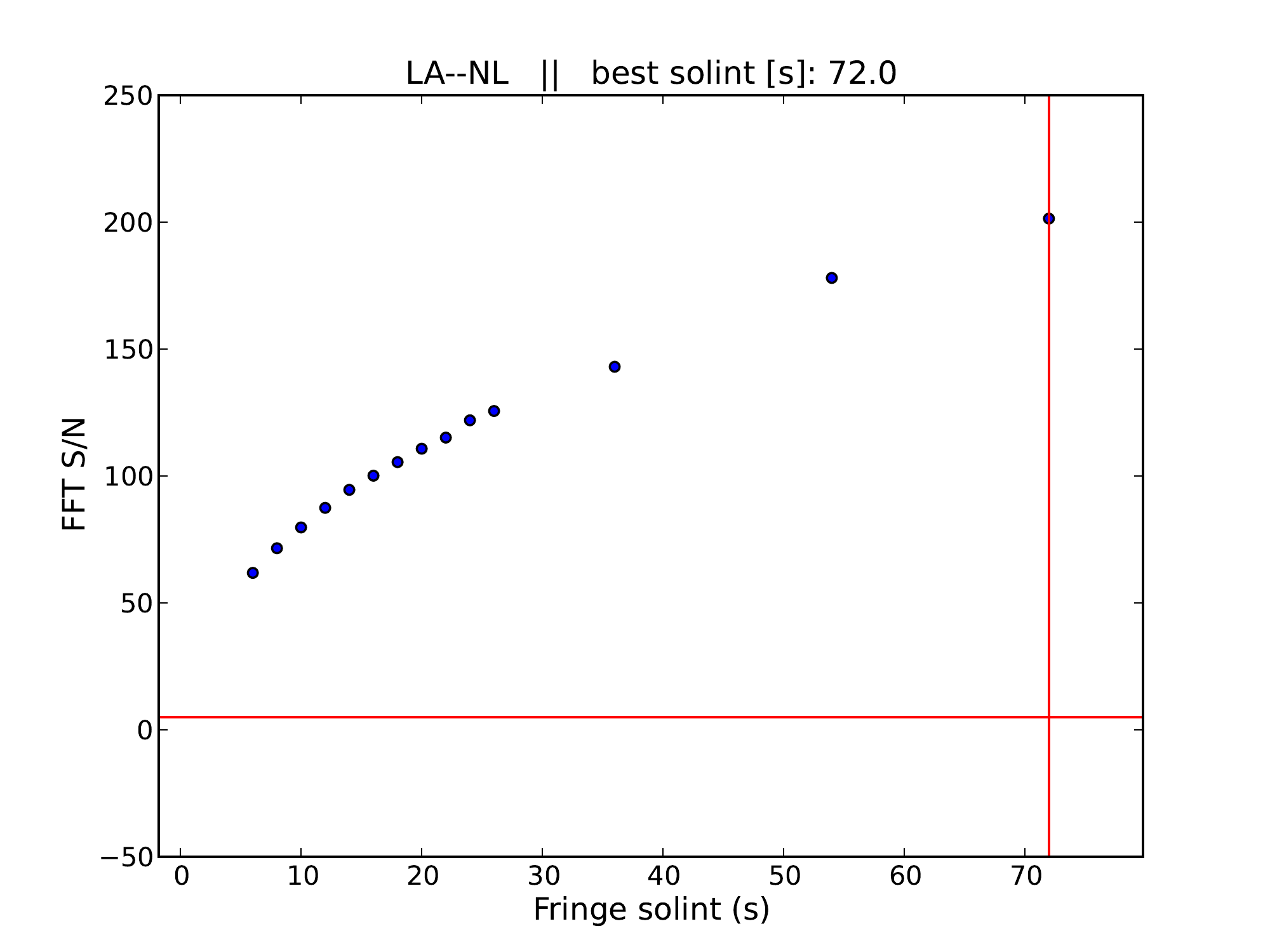}
      \caption{Example plots of a fringe fit solution interval optimization search based on 7mm VLBA data (Section~\ref{verification}).
      Los Alamos (LA) is chosen as the reference station for the scan shown. The blue points show the 
      FFT S/N as a function of solution interval, the horizontal red line indicates the S/N cutoff of 5 on each baseline,
      and the vertical red line shows the 
      smallest solution interval of 72 seconds needed to reach the S/N cutoff on each baseline for this scan.
      The top plot shows the baseline to the Saint Croix (SC) station, which has driven the solution interval
      to higher values until a detection was made when we integrated for 72 seconds.
      The bottom plot shows data from the same scan for the baseline to the North Liberty (NL)
      station, where the source was detected for each solution interval, yielding the expected increase in S/N 
      with the square root of the solution interval.
              }
         \label{solintsearchplots}
   \end{figure}

The solution interval on which the antenna-based corrections are performed is the essential parameter for the phase 
calibration of VLBI data. The shortest timescales on which baseline phases vary are driven by the coherence times
of the atmospheres above the two stations. However, the S/N is often not sufficiently high for the fringe fitter to find solutions
on these short timescales. Therefore, the optimal fringe fit solution interval should be as close as possible to the phase
variation timescales while being high enough to allow for reliable fringe detections for all stations. It follows that 
these intervals strongly depend on the observed source (flux density and structure), antenna sensitivity, weather, and the
observing frequency.

\pipe{} determines phase calibration solution intervals
for each scan within an open search range, based on the observing frequency and array sensitivity. The
default search ranges are given in the online documentation.
The search algorithm uses the S/N from the fast Fourier transform (FFT) stage of the fringe fit process as a metric (Appendix~\ref{ffbasics}).
The FFTs are determined quickly\footnote{
FFTs are limited only by disk I/O, but the least-squares algorithm for a full global fringe fit requires significant CPU time.}
, and an S/N value of 5.5 can be taken as indicator for a solid detection.
The search stops at the smallest solution interval, where the median S/N on each baseline is above the detection threshold,
as shown in Fig.~\ref{solintsearchplots}.
Baselines where the S/N is below the detection threshold are discarded from the search.

In the millimeter regime, short solution intervals are used to calibrate for the atmosphere-induced phase fluctuations.
Here, it may be desirable to obtain fringe solutions for different antennas on different timescales. Typically, this becomes necessary
when some baselines are much more sensitive than others. In this case, fringe solutions
within the coherence time can be obtained for some stations, while much longer integration times may be necessary to obtain detections for 
others. \pipe{} solves this problem with a two-step search in two different ranges of fringe fit solution intervals. The first
search is made on short timescales, and for each scan, any antenna that does not detect the source is recorded. 
In a subsequent search for detections in longer solution intervals, all scans with failed solutions
are again fringe fit.
Detections obtained in the smallest successful solution interval from this second search are used to replace the previously failed solutions.
The default search intervals are given in the online documentation.

The \textit{fringefit} task is used to perform the FFT over the full observing bandwidth for the solution interval search.
Even when instrumental phase and delay offsets are present in the data, the FFT will still have a
decent sensitivity. The optimal solution intervals are first determined for the bright calibrator sources.
These sources are always easily detected, and for high-frequency observations, it is beneficial to calibrate
for atmospheric effects before the calibrator data are used to solve for instrumental effects.
For the science targets, the solution interval search is conducted after all instrumental effects are
solved to obtain more detections in the low S/N regime.

\subsubsection{Coherence calibration for high-frequency observations}
\label{incrcoher}

The first step of fringe fitting typically solves for instrumental phase and delay offsets. 
This single-band fringe fitting is made over
scan durations to maximize the S/N (Section~\ref{instrphdela}).
For high-frequency observations above 50\,GHz (e.g., for the EHT and GMVA), the short coherence times degrade the S/N
significantly over scan durations, which makes it more difficult to obtain robust instrumental calibration solutions.
To overcome this problem, \pipe{} will first solve for phase and rate offsets
on optimized atmospheric calibration timescales that are determined with the method introduced in Sect~\ref{solints}
before the instrumental phase and delay calibration.
This coherence calibration will take out fast intra-scan phase rotations and thereby increase the phase coherence
within scans. With the improved coherence, better instrumental calibration solutions can be obtained.
The coherence calibration is made for the calibrators with the full bandwidth of the observation, that is, 
using data from all spectral windows to maximize the S/N. No delay solutions are applied from this
fringe search; the atmosphere-induced multi-band delays are instead corrected in a later fringe search after the
instrumental phase and delay calibration.
In the cm regime, where the coherence times are much longer, the coherence calibration procedure is not necessary.


\subsubsection{Instrumental phase and delay calibration}
\label{instrphdela}

Depending on the front and back ends of the telescopes, different spectral windows can have different phase and delay offsets.
These data corruptions are mostly constant or vary on very long timescales, meaning that they can be taken out with 
a few observations of bright calibrator sources by fringe fitting the data from each spectral window separately.

\pipe{} will integrate over each scan of the brightest calibrators for this calibration step.
No distinct time-dependent phase drifts should occur for different spectral windows.
Consequently, rate solutions are obtained from the scan-based multi-band fringe fitting step (Section~\ref{mbcal}) and only phase and delay solutions from this instrumental calibration procedure are applied to the data; the single-band solutions for the rates are zeroed.
The instrumental phase and delay solutions are clipped with a high S/N cutoff of 10 by default, 
to ensure a very low false-detection probability.
For most VLBI experiments, instrumental effects are sufficiently stable and \pipe{} will
apply a single solution per spectral window and per station to an entire observation.
These solutions are determined for each station $m$ as follows:
\begin{enumerate}
\item For each scan $s$, the average fringe fit S/N across all $N_\mathrm{spw}$
spectral windows is computed as
\begin{equation}
\overline{\mathrm{S/N}}_m(s) = \frac{1}{N_\mathrm{spw}}\sum_{w=1}^{N_\mathrm{spw}} \mathcal{R}_m(s, w) \;,
\end{equation}
where $\mathcal{R}_m(s, w)$ denotes the fringe S/N of a specific spectral window $w$.
\item A single scan $s_\mathrm{max}$, with the highest $\overline{\mathrm{S/N}}_m$ is selected:
\begin{equation}
\overline{\mathrm{S/N}}_m(s_\mathrm{max}) \geq \overline{\mathrm{S/N}}_m(s) \;  \forall s \;.
\end{equation}
\item The solutions from scan $s_\mathrm{max}$ for all spectral windows
are applied to the data. The solutions from all other scans are discarded.
\end{enumerate}

In some cases, small drifts can occur in the electronics for long observations, or equipment
along the signal path is restarted or replaced, which can cause sudden changes in the instrument.
In these cases, time-dependent instrumental
offsets will be present in the data. \pipe{} is able handle these effects by keeping
the instrumental fringe solutions from all scans $s$ and antennas $m$ where \mbox{$\mathcal{R}_m(s, w) >10 \; \forall w$},
instead of using solutions from a single scan only (which is the default mode described above).
In this special mode, the time-dependent instrumental offsets are corrected by interpolating
the remaining fringe solutions across all scans. Linear interpolation is used for smooth drifts, while
abrupt changes are captured by a nearest-neighbor interpolation.

\subsubsection{Multi-band phase calibration of calibrator sources}
\label{mbcal}

After the instrumental phase and delay calibration, the visibility phases are coherent across spectral windows.
This allows for an integration over the whole frequency band for an increased S/N to calibrate atmospheric effects.
The solutions intervals are determined following the method introduced in Section~\ref{solints}.
A slightly lower S/N cutoff (around 5) can be employed for this global fringe fitting step compared to
the threshold on which the optimal solution intervals are based. This ensures that solutions on short timescales
are not flagged within scans when the S/N fluctuates around the chosen S/N cutoff for detections.
Delay and rate windows (not necessarily centered on zero)
can be used for the FFT fringe search to reduce the probability of false detection in the low S/N regime.
No windows are put in place by default because the sizes of reasonable search ranges are very different
for different observations. The user should inspect the fringe solutions obtained by \pipe,{} and if
there are outliers, repeat the fringe fitting step with windows that exclude the misdetections
or flag the outliers in the calibration table.
Because the atmosphere-induced multi-band delays are continuous functions, the solutions can be smoothed 
in time within scans to remove imperfect detections. First, delay solutions outside of the specified search windows are replaced
by the median delay solution value of the whole scan.\footnote{The starting point for the fringe fit globalization step
is constrained to lie within the search widows. However, solutions can wander outside of that range during
the unconstrained least-squares refinement.} Then, a median sliding window with a width of 60 seconds by default
is applied to the delay solutions, leaving the fast phase and rate solutions untouched. Depending on the stability of
the multi-band delays within scans, different smoothing times can be set in the input files.
To ensure constant phase offsets between the parallel-hand correlations of the different polarization signal
paths, the same rate solutions are applied to all polarizations. They are formed from the average rates of the
solutions from the individual polarizations, weighted by S/N.

The multi-band fringe fit is first made for the calibrator sources . This can be performed in a single pass
because the calibrators are easily detected, even on short timescales.
The weaker science targets are calibrated after the complex bandpass and instrumental polarization
effects have been corrected.

\subsubsection{Complex bandpass calibration}
\label{bpass}

When the calibrator source phase, rate, and delay offsets are calibrated, yielding coherent visibility phases in
frequency and time, the S/N should be high enough to solve for the complex bandpass.
While the scalar bandpass calibration (Section~\ref{scalar-bpass}) can only solve for amplitudes,
the complex bandpass calibration makes use of the cross-correlations and can therefore solve for amplitude and phase
variations introduced by the passband filter of each station within each spectral window.
\pipe{} uses the CASA \textit{bandpass} task to solve for the complex bandpass.
Single solutions for each antenna, receiver, and spectral window over entire experiments are
obtained based on the combined data of all scans on the specified 
bandpass calibrators. If the S/N is good enough, per-channel solutions can be obtained.
Otherwise, polynomials should be fitted. If a prior scalar bandpass calibration was done, the amplitude solutions
from the complex bandpass calibration are set to unity. 
If the amplitudes are to be solved for here,
flat spectrum sources should be chosen as bandpass calibrators (this is especially important
for wide bandwidths) or an a priori source model encompassing
the frequency structure must be supplied to \pipe{}.
If the S/N is too low or not enough data on bright calibrators are available, the complex bandpass
calibration should be skipped.

\subsubsection{Phase calibration of science targets}
\label{mbphsci}

\pipe{} will solve for instrumental phase and delay offsets, the complex bandpass, and polarization leakage solutions (Section~\ref{polcalsect}).
After all these instrumental data corruptions are removed, the phases of the weaker science targets are calibrated.
In this step, all previous calibration tables are applied.
These are the sampler corrections, the amplitude calibration, the amplitude bandpass, the phase bandpass,
the instrumental phase and delay calibration, the polarization solutions, and calibrator multi-band fringe solutions in the case
of phase-referencing (Appendix~\ref{phrefastrom}).

Three different science target phase calibration paths are implemented in \pipe{}:
\begin{enumerate}
\item If phase-referencing is disabled, a first multi-band fringe search is made over long
integration times to solve for the bulk rate and delay offsets with open search
windows for each scan. This will determine whether the source can be detected;
fringe non-detections will flag data.
Then, intra-scan atmospheric effects are 
corrected for in a second multi-band fringe fit with narrow search windows following the
method presented in Section~\ref{mbcal}.
This is the default option for high-frequency observations.
\item If phase-referencing is enabled and no fringe search on the science targets
for residuals is to be done, phases, delays, and rates are calibrated
using only the phase-referencing calibrator solutions.
All valid data of the science targets will be kept.
This is the default option for low-frequency observations if the
science targets are very weak or for astrometry experiments.
\item If phase-referencing and a search for residuals on the science targets 
are enabled, the calibration is done with the same two-step fringe fitting approach
as in option 1 while also applying phase-referencing calibrator solutions
on-the-fly. Here, very narrow search windows can be used.
The goal is to solve for residual phase, delay, and rate offsets
that are not captured by the phase-referencing.
This is the default option for non-astrometry experiments in the cm regime with
strong enough science targets.
\end{enumerate}

\subsection{Polarization calibration}
\label{polcalsect}

All calibration solutions described so far are obtained independently for the different polarization
signal paths (RCP and LCP for circular feeds).
To calibrate the cross-hand correlations, strong, polarized sources must be observed over a wide range of feed rotation angles.
This enables imaging of all four Stokes parameters. For circular feeds, the Stokes parameters for total intensity $I$,
linear polarization $Q$ and $U$, and circular polarization $V$, are formed as

\begin{eqnarray}
\label{stokes}
I & = & \frac{1}{2} \left( \mathrm{RR} + \mathrm{LL} \right) \\
Q & = &\frac{1}{2} \left( \mathrm{RL} + \mathrm{LR} \right) \\
U & = &\frac{i}{2} \left( \mathrm{LR} - \mathrm{RL} \right) \\
V & = &\frac{1}{2} \left( \mathrm{RR} - \mathrm{LL} \right) \; ,
\end{eqnarray}
with $i=\sqrt{-1}$.

Feed impurities are typically constant over experiments. 
Therefore, constant solutions are obtained for the polarization
calibration by combining the data from all polarization calibrator scans.
The CASA \textit{gaincal} task with the type `KCROSS' setting is used to solve for
cross-hand delays and the \textit{polcal} task with the type `Xf' is used to solve
for cross-hand phases.
Depending on the S/N of the cross-hand visibilities on the polarization
calibrator sources and the instrumental polarization structure, cross-hand phases can either
be solved per spectral window, per individual frequency channel, or by binning groups of
frequency channels together within each spectral window. The cross-hand delays are always
solved per spectral window.
Both tasks determine solutions under the assumption of a
point source model with 100\% Stokes $Q$ flux here.
This assumption will affect
the absolute cross-hand phase that determines the electric vector position angle (EVPA)
orientation across the source: if the source is imaged,
adding an arbitrary phase will rotate the EVPA $\Phi$ of each linear polarization vector
by the same amount. The absolute EVPA
can be calibrated using connected-element interferometric or single-dish observations under the assumption
that the field orientation remains the same on VLBI scales.
The EVPA follows from the Stokes $Q$ and $U$ fluxes as
\begin{equation}
\label{evpa}
\Phi = \frac{1}{2}\tan^{-1}\left( \frac{U}{Q} \right)
.\end{equation}

Secondly, the leakage,  or D-terms \citep{Conway1969},
between the two polarizations of each receiver are corrected.
Some power received for one polarization will leak into the signal path of the other one and vice versa,
causing additional amplitude and phase errors in the cross-hand visibilities.
Depending on the S/N and instrumental frequency response, \pipe{} can obtain leakage solutions for the whole frequency band,
per spectral window, or per group of frequency channels within each spectral window.
It can be assumed that extragalactic synchrotron sources have a negligible fraction
of circular polarization\footnote{
For a typical observation that includes multiple calibrator sources, this assumption is easily tested.}
and that the parallel-hand correlations are perfectly calibrated, including corrections for the feed rotation angle $\chi$
of the stations (Appendix~\ref{mounttypes}).
In this case, 
small D-terms that can be modeled by a first-order approximation
for circular feeds on the $m$-$n$ baseline can be written as \citep{Leppanen1995}
\begin{eqnarray}
\label{dterms}
\mathrm{RL}_{mn}^\mathrm{obs} & = & \mathrm{RL}_{mn}^\mathrm{true} + \left( D^R_m e^{2 i \chi_m} + \left(D^L_n\right)^* e^{2 i \chi_n} \right) I\\
\mathrm{LR}_{mn}^\mathrm{obs} & = & \mathrm{LR}_{mn}^\mathrm{true} + \left( D^L_m e^{-2 i \chi_m} + \left(D^R_n\right)^* e^{-2 i \chi_n} \right) I \; .
\end{eqnarray}
Here, $D$ are the leakage terms, with a superscript indicating the polarization, and subscript indicating the antenna.
The effect of the D-terms is a rotation in the complex plane by twice the feed rotation angle, which makes
this instrumental effect discernible from the true polarization of the source, which is usually constant during observations.
The complex D-terms are estimated with the CASA \textit{polcal} task.
An S/N-weighted average of the individual calibration solutions will be formed
when more than one polarization calibrator is used.
It should be noted that while the D-terms have to be determined after the calibration of cross-hand phase and delay offsets,
CASA will always apply the solved calibration tables in the signal path order, that is, leakage before instrumental delays
and phases.\footnote{Within CASA and \pipe{}, iterative calibration solutions are easily obtained, which
allows for incremental corrections to the D-terms and cross-hand phase and delay offsets.
Normally, this is not necessary.}
This approach is similar to the AIPS PCAL task, and
it requires a good total intensity source model. However, for strongly resolved sources with varying polarization
structure along the different spatial components, \textit{polcal} will fail to determine accurate D-terms.
To address this issue, a task similar to AIPS LPCAL will be added to the pipeline in the future (Section \ref{futurefeatures}).

\subsection{Post-processing and application of calibration tables}
\label{calsolpostprocapp}
The solution tables obtained from each calibration steps described in this section
can be post-processed with user-defined median or mean filter smoothing,
and different options for time and frequency interpolation schemes offered by CASA
(nearest, linear, cubic, or spline) can be used for the application of the solutions.
The default linear interpolation should be the best option in most cases.

\subsection{Diagnostics}
\label{diagsect}

By default, \pipe{} stores many diagnostics from every run in a dedicated folder within the working directory:
\begin{itemize}
\item Plots of calibration solutions from each calibration table. Where applicable, plots will be made on a per-source basis.
\item Solution interval searches as shown in Fig.~\ref{solintsearchplots}.
\item Receiver temperature fits as shown in Fig.~\ref{tsysstar}.
\item Plots of visibilities made with the jplotter\footnote{\url{https://github.com/haavee/jiveplot}.} software.
These plots can be made for the raw
visibilities and once again after applying the calibration solutions; a comparison of these plots serves as a good metric
for the performance of the pipeline. The visibility plots are made for a number of scans that are selected 
from the start, middle, and end of the experiment based on the number of baselines present. The plots show
visibility phases and amplitudes, as a function of time (frequency averaged) and as a function of frequency (time averaged).
\item Lists of all applied flags and an overview of flagged data percentages per station.
\item Text files, which show fringe detections for all scans across the array.
\item CASA log files containing detailed information about every step of the pipeline.
\item Copies of the used command line arguments and configuration files.
\end{itemize}
These logging procedures, together with the use
of a single set of input files that determine the whole pipeline calibration process, makes the results fully reproducible.
Examples of diagnostic plots from a full calibration run are presented in Appendix~\ref{m87calib}.

\section{Imaging and self-calibration}
\label{imagscal}

   \begin{figure*}[h]
   \centering
   \includegraphics[width=0.85\textwidth]{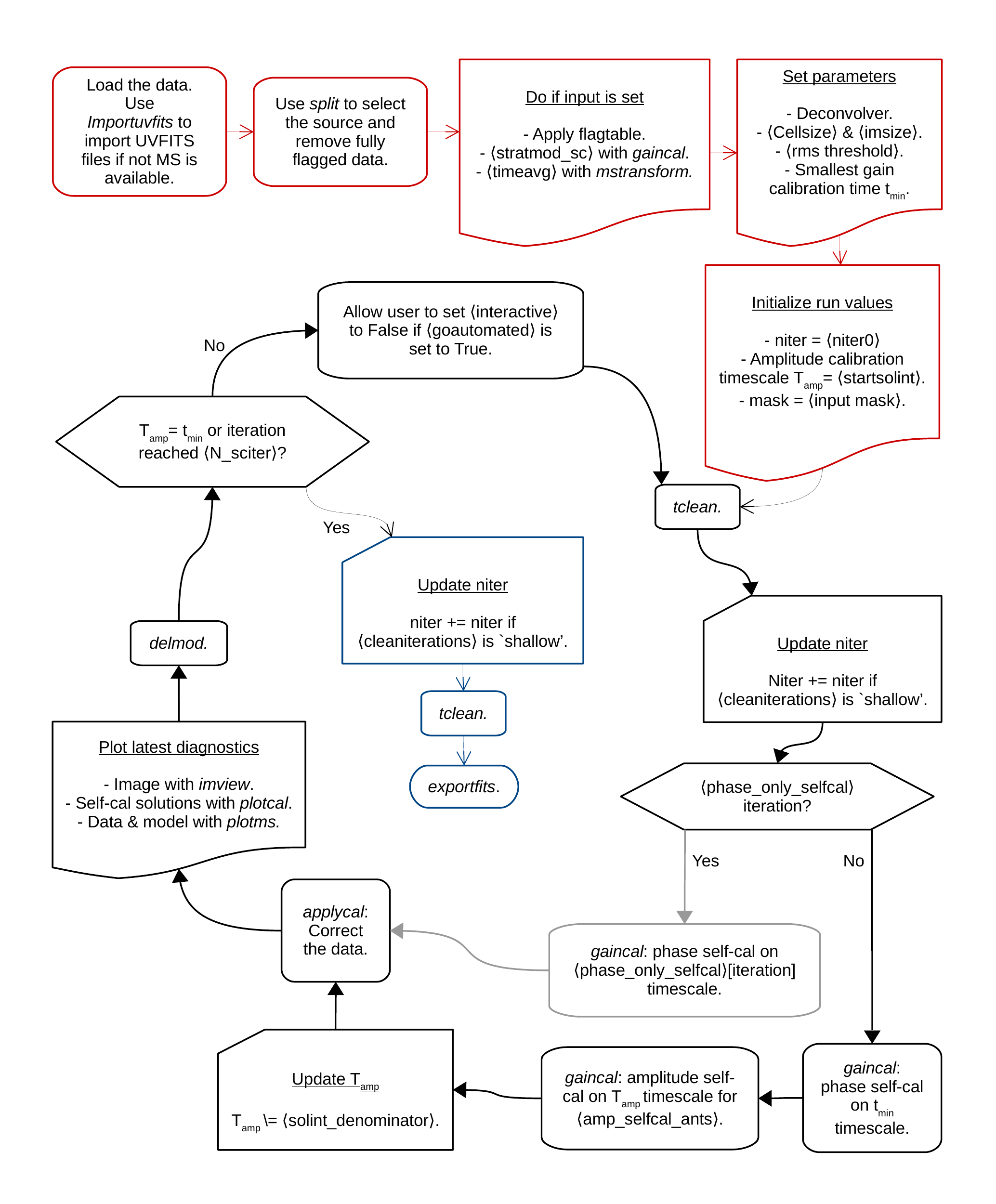}
   \caption{Overview of the \pipe{} imager performing \textit{tclean} and
            self-calibration iterations.
            CASA tasks used by the pipeline are written in italics,
            and input parameters are written inside angle brackets.
            The solutions from each self-calibration table are incremental
            with respect to all previous solutions. 
            Interpolated values from good self-calibration results
            are used instead of failed (flagged) solutions unless an
            antenna is fully flagged.
            Optionally, only phase self-calibration can be performed
            or a single image can be made without any self-calibration.
            Within each \textit{tclean} operation, the mask is updated between
            CLEAN cycles, by CASA if `auto-multithresh' is enabled and/or
            by the user in interactive mode.
            }
              \label{image_flowchart}
    \end{figure*}

\begin{figure*}[h]
\centering
  \begin{minipage}[b]{0.5\linewidth}
    \centering
    \includegraphics[width=.85\linewidth]{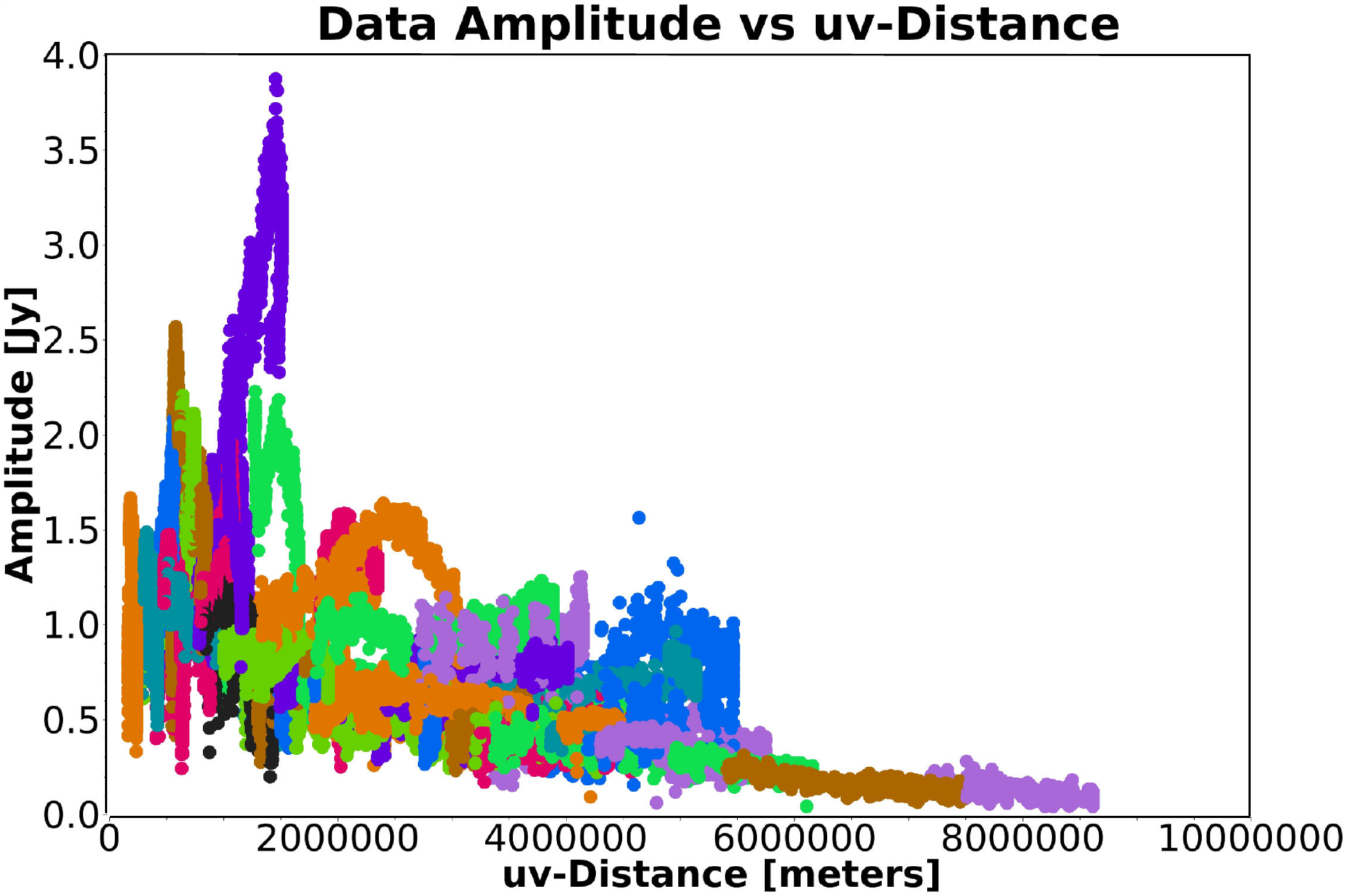}
  \end{minipage}
  \begin{minipage}[b]{0.5\linewidth}
    \centering
    \includegraphics[width=.85\linewidth]{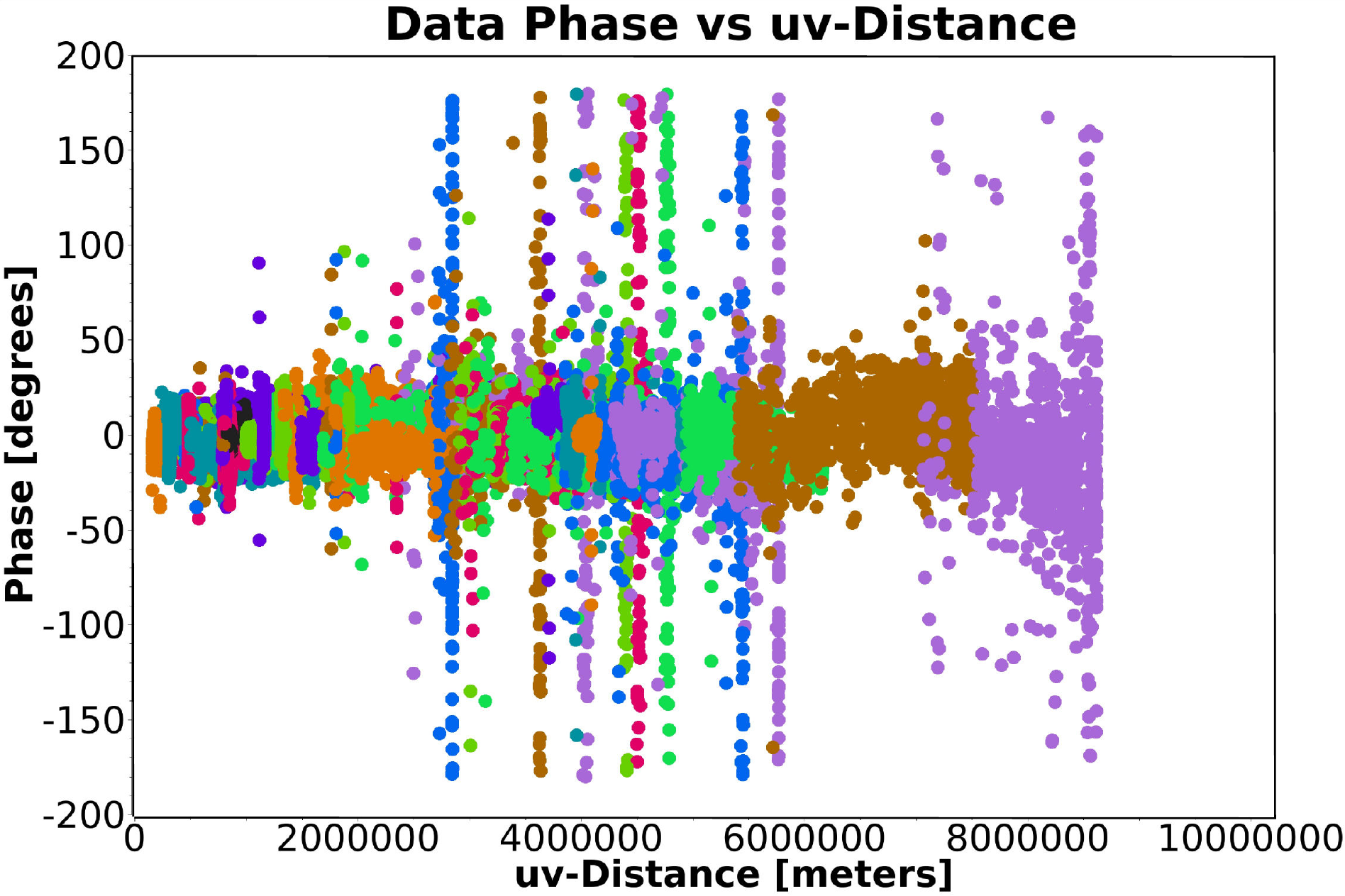}  
  \end{minipage} 
  \begin{minipage}[b]{0.5\linewidth}
    \centering
    \includegraphics[width=.85\linewidth]{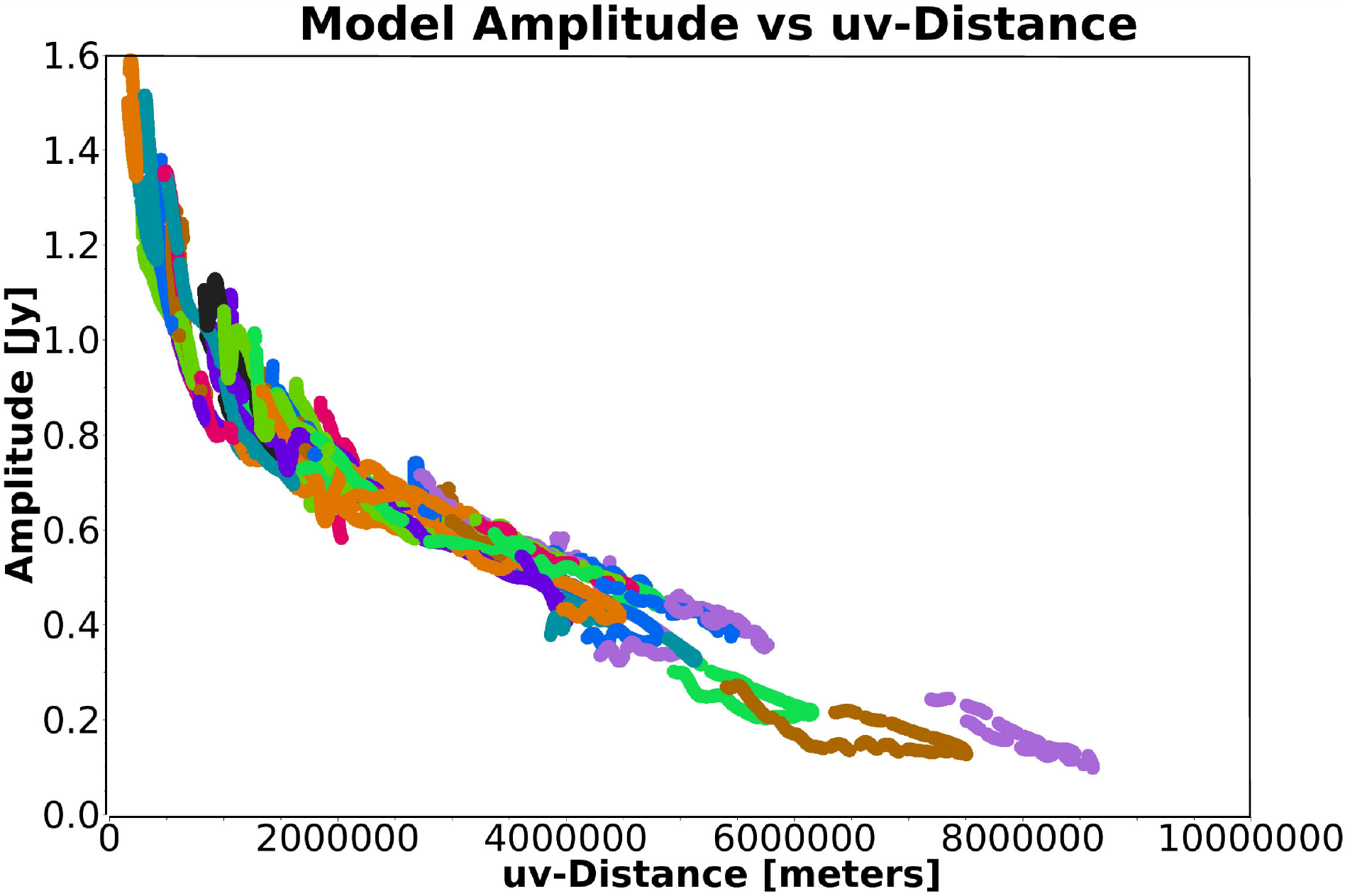}  
  \end{minipage}
  \begin{minipage}[b]{0.5\linewidth}
    \centering
    \includegraphics[width=.85\linewidth]{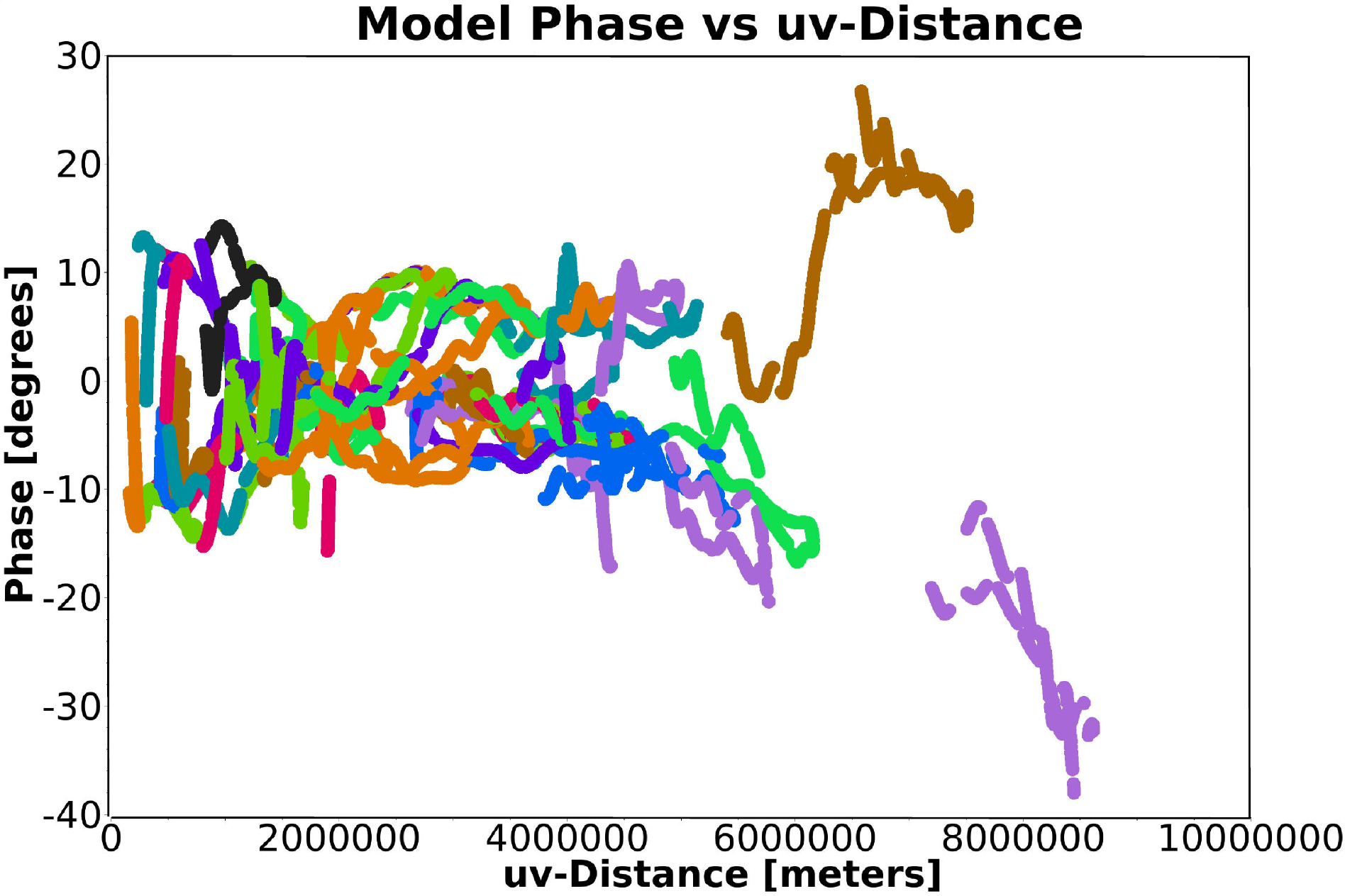}  
  \end{minipage} 
    \begin{minipage}[b]{0.5\linewidth}
    \centering
    \includegraphics[width=.85\linewidth]{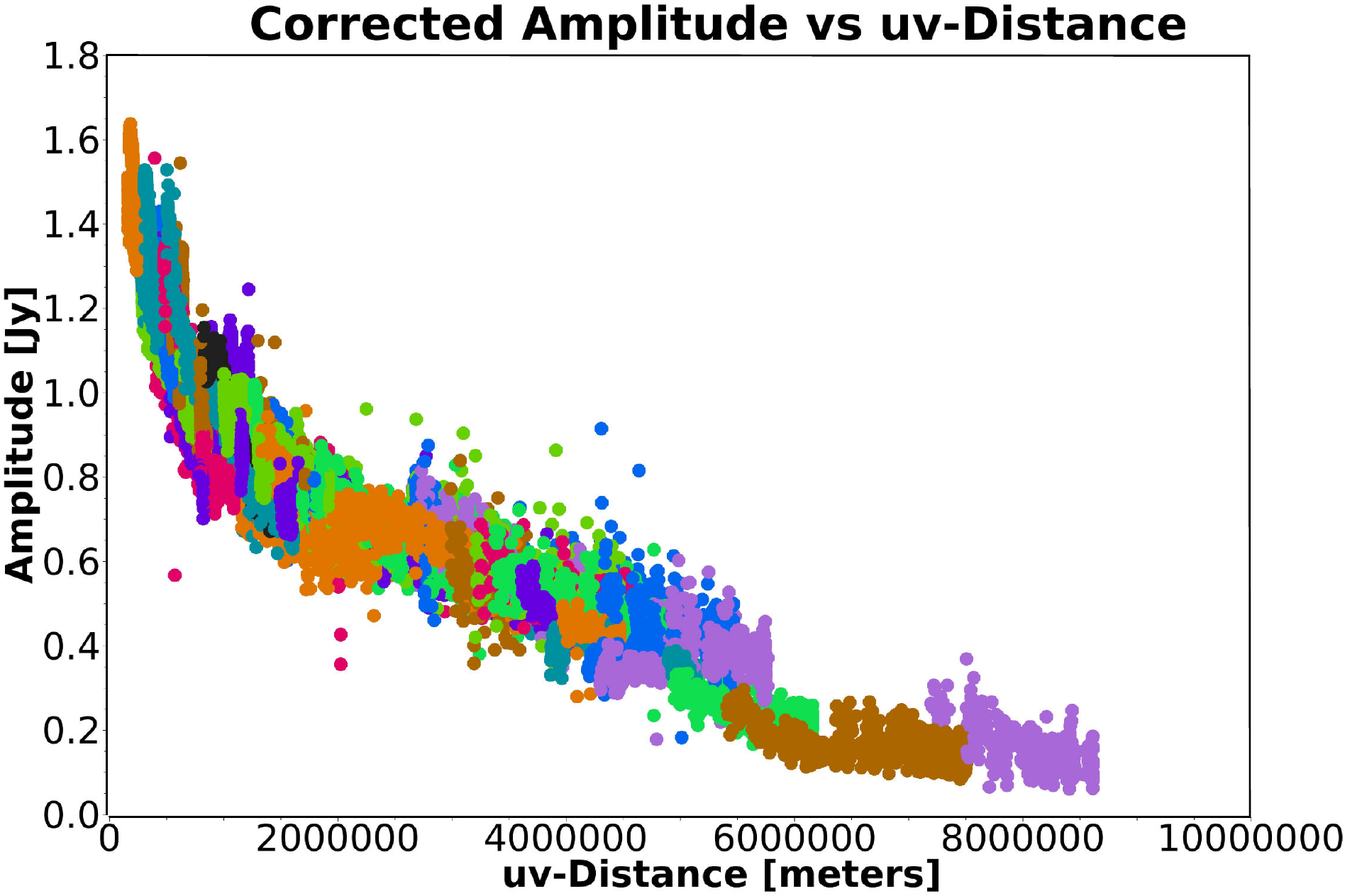}  
  \end{minipage}
  \begin{minipage}[b]{0.5\linewidth}
    \centering
    \includegraphics[width=.85\linewidth]{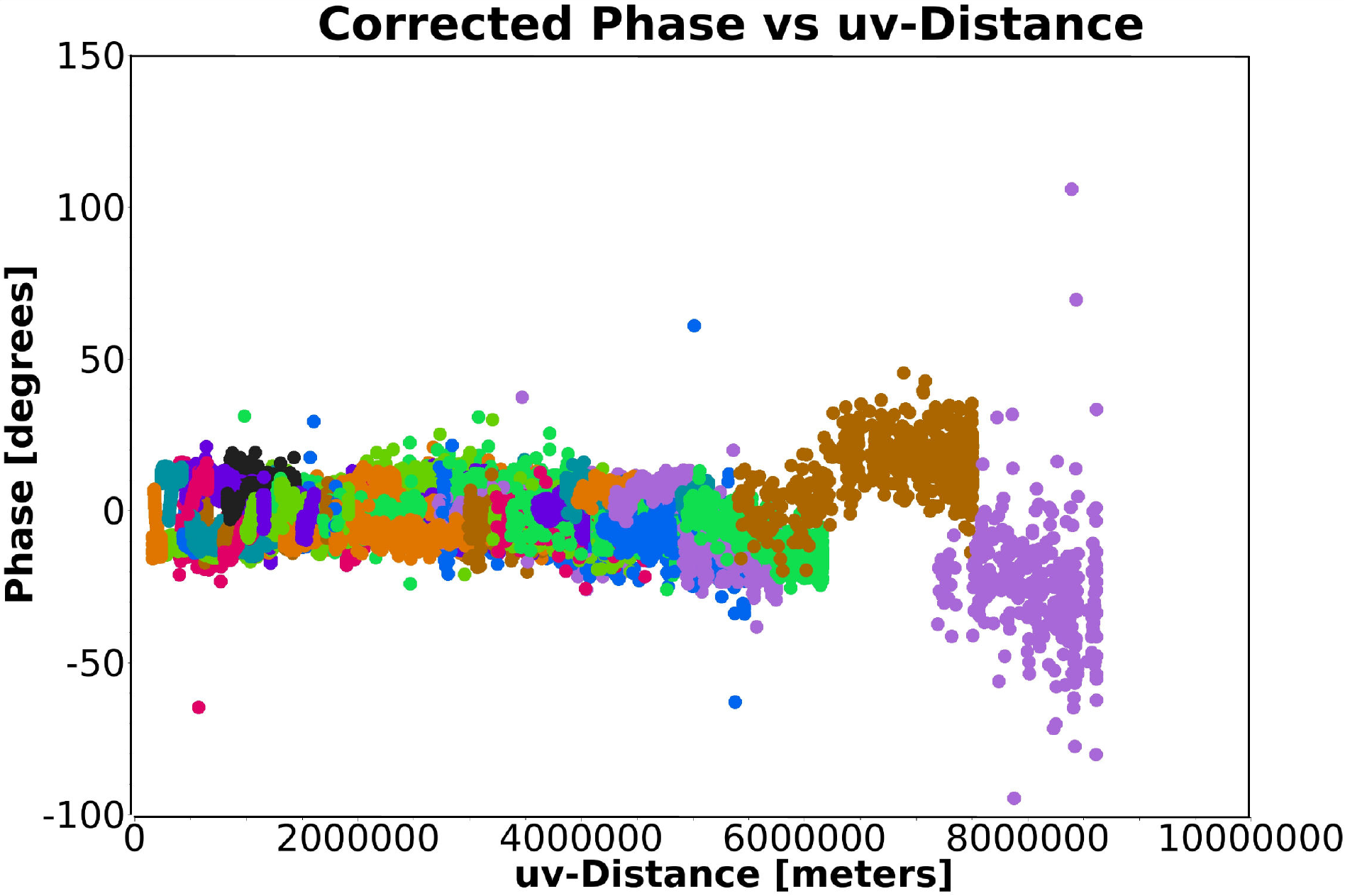}  
  \end{minipage}
\caption{M87 amplitudes on the left and phases on the right as a function of baseline length (u-v distance),
         color-coded by baselines.
         The top panels show visibilities after the model-agnostic \pipe{} calibration (before
            self-calibration). 
            The middle panel presents the model data from the final image of the 
            imaging plus self-calibration cycles, shown in Fig.~\ref{m87imaging}.
            The bottom panel depicts the data after the last round of self-calibration.
            In the model-agnostic calibration, 
            anomalously high amplitudes are due to station gain errors;
            the system temperature measurements of the Fort Davis and Owens Valley stations were doubtful.
            Large residual phase trends are related to low S/N measurements, where long fringe fit solution intervals
            are chosen by \pipe{}. These amplitude and phase trends are both corrected for with the
            iterative self-calibration.
        }
                 \label{radplots}
\end{figure*}

The model-agnostic calibration described in Section~\ref{picalib} is typically done without a priori knowledge about
the source model. This limits the amplitude calibration to the measured SEFDs, which can have large
uncertainties. Values for the DPFUs and gain curves are affected by measurement errors, system temperature
variations during scans are often not captured, phasing efficiencies for phased interferometers
have uncertainties, and off-focus as well as off-pointing amplitude losses are not captured in most cases.
With the default assumption of a point source in the model-agnostic calibration, the phase calibration
will naturally steer baseline phases to the reference station toward zero (Appendix.~\ref{ffbasics}).
When the correct source model is known, the phases can be calibrated toward the true values and on shorter
(atmospheric) timescales.

The basic principle of self-calibration is to image the data to obtain a source model
and to use that model to derive station-based amplitude and phase gain solutions,
correcting for the aforementioned shortcomings in the model-agnostic calibration
\citep{Readhead1978, Pearson1984}.
For CLEAN-based imaging algorithms \citep{Hogbom1974}, this is often an iterative process.
As the source model improves by recovering weaker emission features, better
self-calibration solutions are obtained, which will in turn improve the 
imaging. This process should converge to a final image after a finite number
of self-calibration iterations on increasingly shorter solution intervals.
It should be noted that phase self-calibration degrades the astrometric 
accuracy of phase-referencing experiments.

Within \pipe{}, imaging and self-calibration capabilities are added as an interactive feature.
It is a single module, independent from the rest of the pipeline, that can be used in an
interactive CASA session to image sources from any MS or UVFITS file.
The CASA \textit{tclean} function is used as an MPI-parallelized, multi-scale, multi-frequency synthesis 
modern CLEAN imaging algorithm. The robustness scheme can be used to weight the data \citep{Briggs1995}.
The \pipe{} imager will perform iterations of imaging and incremental phase plus amplitude self-calibration steps. The CASA \textit{gaincal} task is used for the self-calibration.
Fig.~\ref{image_flowchart} provides an overview of the algorithm steps, and the most important
input parameters are given in the online documentation.

To avoid cleaning too deeply and thereby picking up source components from the noise,
the maximum number of clean iterations is set to a low number for the first few 
iterations. As the data improve with every self-calibration iteration, the
maximum number of clean iterations is gradually increased. 
Additionally, the cleaning will stop earlier if the peak flux
of the residual image is lower than the theoretically achievable point-source sensitivity of the array
or (optionally) when a $3 \sigma_n$ stopping threshold has been reached, with $\sigma_n$ given as
1.4826 times the median absolute deviation in the residual image.

For the self-calibration, S/N cutoffs of 3 for the phases and 5 for the amplitudes are
employed to avoid corrupting the data
by an erroneous source model. Failed solutions are replaced by
interpolating over the good solutions, except for the final self-calibration
iteration. This ensures that data without sufficient
quality for self-calibration will be excluded from the final image.
First, the self-calibration is made for visibility phases,
normally starting with a a point-source model, which allows for a subsequent 
time-averaging with reduced coherence losses and results in a better starting point of a bright component
at the phase center for the imaging.
Next, as the CLEAN model is constructed, a few phase-only self-calibration steps
are preformed on increasingly shorter solution intervals
to first recover the basic source structure.
Then, the self-calibration is used to adjust amplitude gains on
long timescales at first to avoid freezing-in uncertain source components. 
These solution times are then lowered by a factor of 2 for every iteration as the model improves and
definite source components are obtained.
If necessary, data from certain u-v ranges can be excluded from the amplitude self-calibration,
for instance, long baselines with low S/N.

The imaging function can be run entirely interactively, where CLEAN boxes can
be placed and updated for each major cycle of each imaging iteration.
When a final set of CLEAN boxes is obtained, the algorithm can be set to
run automatically for all remaining iterations. 
Alternatively, the imaging can be fully automated when a set of CLEAN masks is supplied
as input or when the \textit{tclean} auto-masking options are used.
This is done with the CASA `auto-multithresh' algorithm.
The auto-masking
algorithm will try to determine regions with real emission primarily based on noise statistics,
and side lobes in the image. A detailed description of the algorithm can be found online in the official CASA documentation.\footnote{
\url{https://casa.nrao.edu/casadocs-devel/stable/imaging/synthesis-imaging/masks-for-deconvolution}
}
The default auto-masking parameters set in \pipe{} should produce reasonable masks for compact sources,
but if faint extended emission has to be recovered, the
conservative auto-masking approach may not be sufficient. In this case, parameter tweaking
or interactive user interaction on top of the auto-masking is required.

The image convergence is easily tracked because
plots of all self-calibration solutions, the model, the data after each self-calibration,
and the images (beam-convolved model plus residuals) of each cycle are made by default.

\section{Science test case: 7mm VLBA observations of the AGN jet in M87}
\label{verification}

As a demonstration of the \pipe{} calibration and imaging capabilities, we
present here the results from an end-to-end processing of a 
representative VLBI dataset. For this purpose, we have selected
continuum observations of the central AGN in M87
conducted with the VLBA at 7\,mm (43\,GHz).
The chosen dataset is an eight-hour-long track on M87, including
3C279 and OJ287 as calibrators, with a bandwidth 
of 256\,MHz that was distributed over two spectral windows with 256 channels each
(PI: R. Craig Walker, project code: BW0106).
All VLBA stations participated in this experiment.

A description of the data calibration with \pipe{} and AIPS (to create a benchmark dataset) is given in §\,\ref{7mmcalib}.
Imaging and self-calibration steps are descried in §\,\ref{7mmimagingandsc}, and
the imaging results are presented in §\,\ref{7mmresults}.

\subsection{Model-agnostic calibration}
\label{7mmcalib}

The data were blindly calibrated with \pipe{} v1.0.0 in a single pass using the pipeline default
calibration parameters for high-frequency VLBA observations. A priori information
was used to flag poor data and to calibrate the amplitudes based on system temperatures,
DPFUs, and gain curves. Auto-correlations were used to 
correct for erroneous sampler thresholds of station recorders and to 
calibrate the amplitude bandpasses after edge channels were flagged.
Instrumental single-band delay and phase offsets together with phase bandpasses
were calibrated using the bright calibrators OJ287 and 3C279. Optimal solution
intervals for the fringe fit phase calibration were determined based on the S/N
of each scan.

Independently, the data were manually calibrated in AIPS, following the standard
recipe outlined in Appendix C of the AIPS cookbook.
In this process, fringe fitting was performed with a 30\,s timescale for all scans (the adaptive scan-based
solution interval tuning is a unique feature of \pipe{}).
Overall, this integration time yields solid fringe detections while still capturing the atmospheric phase variations. 
The AIPS calibrated visibilities serve as a benchmark dataset to cross-compare results between AIPS and CASA/\pipe{}.
Appendix~\ref{m87calib} shows a selection of \pipe{} calibration solutions from the
different pipeline steps together with AIPS calibration solutions for
a direct comparison.

\subsection{Self-calibration and imaging}
\label{7mmimagingandsc}

Both the \pipe{} and AIPS calibrated datasets were imaged
with the \pipe{} imager using the same default settings,  self-calibration steps, and CLEAN windows (Section~\ref{imagscal}).
Initial phase self-calibration steps were made for 300\,s and 10\,s.
After this, the phases were continuously self-calibrated with 10\,s
solution intervals after each imaging iteration.
Only phase gain solutions with $\mathrm{S/N}>3$ were applied to the data.
The first step of amplitude self-calibration used a two-hour  integration, which was lowered
by a factor of 2 for each imaging iteration.
The final amplitude self-calibration step was made on a 10\,s solution interval.
Only amplitude gain solutions with $\mathrm{S/N}>5$ were applied to the data.

Fig.~\ref{radplots} shows the effect of the self-calibration after the model-agnostic
calibration.
Before self-calibration,
amplitude and phase gain errors  are clearly visible (Fig.~\ref{radplots}, upper panels) 
and baseline phases 
are partially steered toward zero because a point-source model was assumed for the fringe fitting (Fig.~\ref{radplots}, upper right panel).
Station-based gain solutions were applied to the data in several self-calibration iterations, which eventually converged to the best source model image (shown in Fig.~\ref{m87imaging}).
At the end of the self-calibration process, the 
corrected visibilities closely matched the model visibilities (Fig.~\ref{radplots}, lower and middle panels). At uv-distances $>6M\lambda,$ the phases deviate from zero because of the spatial structure of the source. 

Fig.~\ref{m87imaging_comparison} shows a direct comparison
of image reconstructions for the \pipe{} and AIPS calibrated data.
The overall reconstructed jet structure and salient image features
described above are consistent in the two reconstructions.
Imaging the same data in Difmap showed no meaningful differences
with Fig.~\ref{m87imaging_comparison}.

\subsection{Results}
\label{7mmresults}

    \begin{figure*}[h]
\centering
    \includegraphics[width=.85\linewidth]{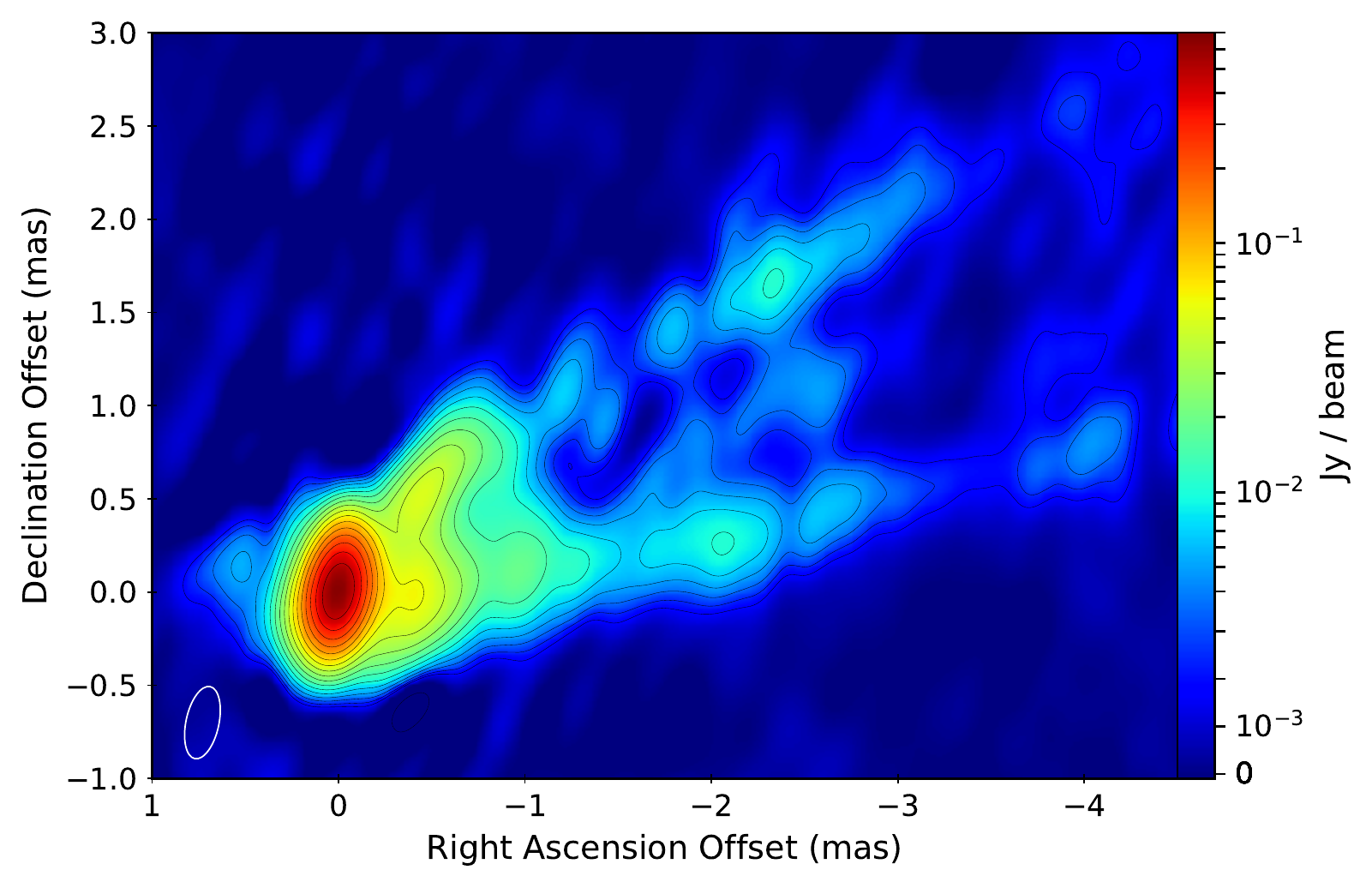}
\caption{CLEAN image reconstruction of the 7mm M87 jet from VLBA data using robust weighting of the data
         ($\mathrm{\textit{robust}} = 0.5$ in \textit{tclean}).
         The data were calibrated and imaged with \pipe{}. 
         The restoring beam (0.37\,x\,0.17\,mas at $-13.3^\circ$), representing the resolving power due to
         the u-v coverage of the data, is shown in bottom left corner.
         The color range displayed is $-10^{-4}\,\mathrm{Jy}/\mathrm{beam}$ to $0.7\,\mathrm{Jy}/\mathrm{beam}$.
         Contours in mJy per beam are drawn for -1.8, 1.8, and 2.5, increasing by factors of $\sqrt{2}$
         from there.
        }
                 \label{m87imaging}
\end{figure*}

\begin{figure*}[h]
\centering
    \hspace{-0.6cm} (\small{a) cal:\pipe{} \& im:\pipe{} (natural)} \hspace{4.1cm} \small{(b) cal:AIPS \& im:\pipe{} (natural)}
  \begin{minipage}[b]{0.5\linewidth}
    \centering
    \includegraphics[width=.85\linewidth]{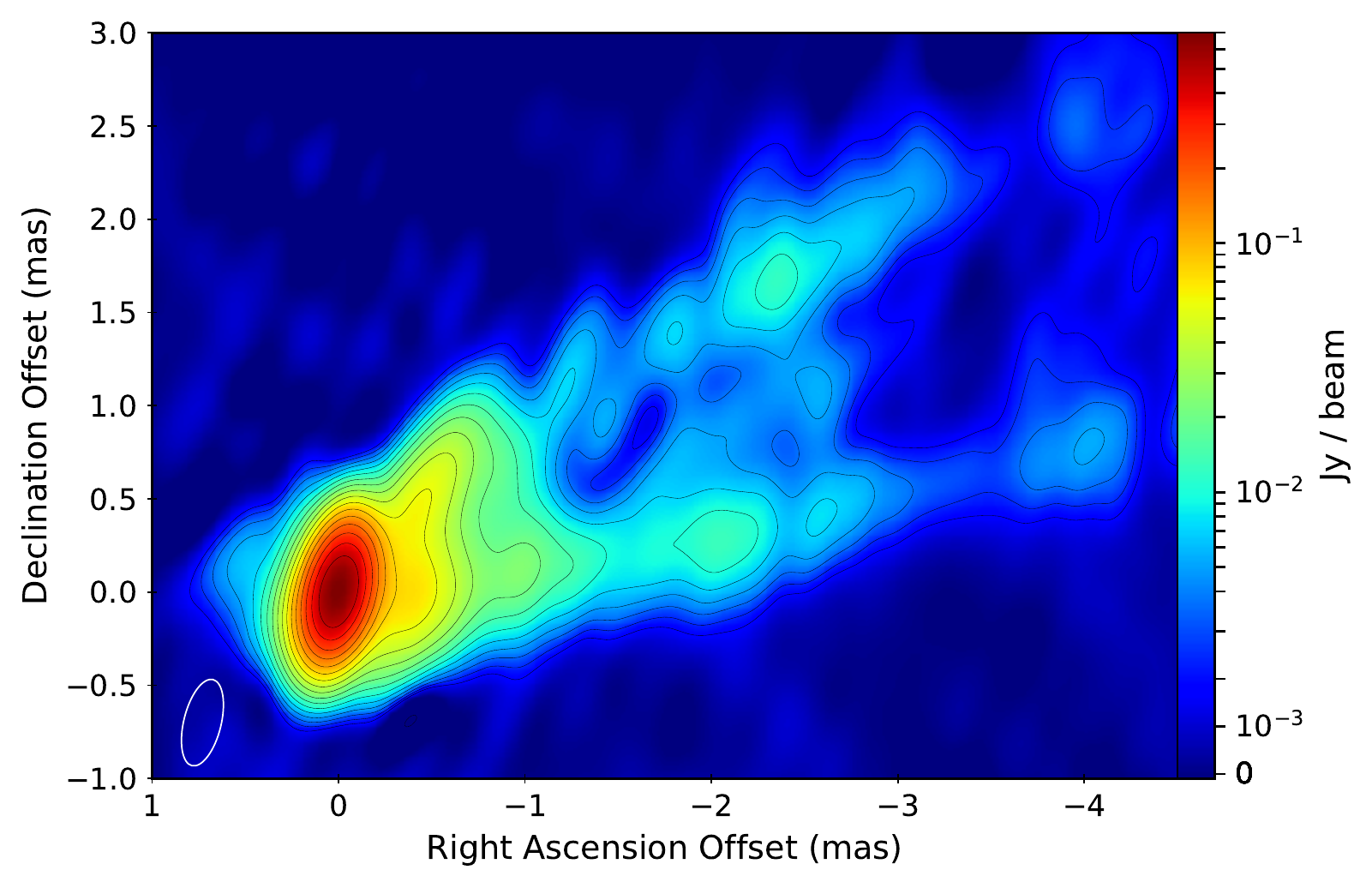}  
  \end{minipage}
  \begin{minipage}[b]{0.5\linewidth}
    \centering
    \includegraphics[width=.85\linewidth]{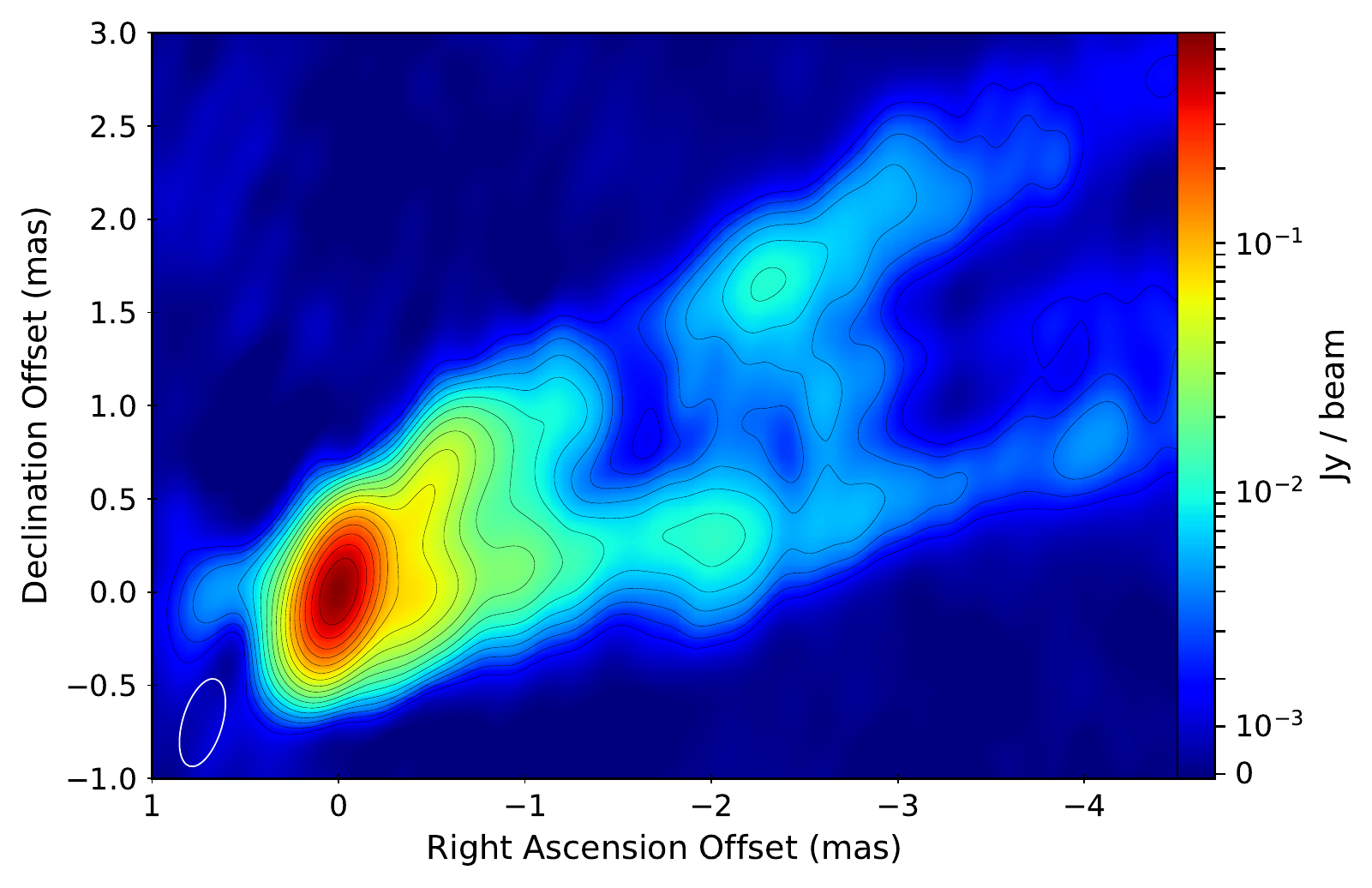}  
  \end{minipage} 
\caption{Comparison of CLEAN image reconstructions of the 7mm M87 jet between 
         \pipe{}-calibrated data in the left panel and
         AIPS-calibrated data in the right panel.
         Both images were produced with the \pipe{} imager using natural weighting.
         The restoring beam, color range, and contour levels are the same as
         in Fig.~\ref{m87imaging} for the two images shown here.
        }
                 \label{m87imaging_comparison}
\end{figure*}

Fig.~\ref{m87imaging} shows the  fiducial image of the science target that was obtained after calibration and imaging with \pipe{}.  
The bright core of the radio jet in M87 is centered at zero offset in the image.
The western part of the jet is much brighter than the
eastern part. This one-sidedness can be explained by Doppler 
boosting of a jet that is inclined with respect to the 
line of sight: the approaching jet is brighter
than the receding counter-jet.
Moreover, the edges of the M87 jet are clearly
brighter than the central jet spine.
This is the result of a rotating sheath of material
that surrounds the jet spine.
The southern
jet arm follows an almost straight trajectory, while the track of the northern part of the jet
at first leaves the jet core at a large opening angle of $\sim50^\circ$ . At a right ascension offset of about
-1 mas, the upper jet arm bends into a recollimated trajectory that extends parallel to the southern arm as seen on larger scales.
The recollimation at that particular position
may be the result of a strong interaction between the jet and the ambient medium surrounding the jet.
This structure is also in accordance with earlier work \citep{Walker2018}.
A weak emission region is present toward the east of the radio core.
The brightness of this feature is about 0.05\,Jy, that is, 10\,\%
of the brightness of the approaching jet close to the radio core. When the weak extension is interpreted as 
a counter-jet, a symmetrical flow profile would be constrained to about half
the speed of light for inclination angles $\lesssim 20^\circ$.

\section{Future features}
\label{futurefeatures}

Because \pipe{} is built upon CASA, it will benefit from the steady CASA core developments.
Notably, these will include enhancements of the MS data structure,
updates of the calibration framework, extensions of the MPI implementation, 
and imaging improvements.

The CASA fringe fitter will be able to fit for a dispersive delay for
low-frequency observations with large fractional bandwidths
in the future. This will be important for arrays such as \mbox{LOFAR} \citep{Haarlem2013} and the SKA.
While an exhaustive fringe search is already implemented in \pipe{} (Appendix~\ref{ffbasics}),
we will also add a baseline-stacking capability to the \textit{fringefit} task itself.

Additional spectral line specific calibration routines are planned to
be included in the next release of \pipe{}: A restriction of the bandpass
calibration to continuum calibrator sources, a compensation for time-variable
Doppler shifts after the bandpass calibration, 
a delay calibration based
only on the continuum calibrators while rate solutions are determined
from the spectral line itself, and spectral line amplitude calibration
based on the auto-correlations and template spectra of lines.

For an improved polarization calibration, a new CASA task called \textit{polsolve}\footnote{
\url{https://code.launchpad.net/casa-poltools}.}
is currently under development and will be integrated into \pipe{}.
This task is able to obtain full nonlinear solutions for the leakage
and more accurate D-terms for extended calibrator sources with varying polarization structure.
After a source has been imaged in total intensity, it will be possible to
decompose a CLEAN model into multiple compact regions
with sufficiently constant polarization structure. D-terms can be solved
for each region separately, and the median leakage will be applied
to the data.
The CASA \textit{tfcrop} algorithm will be explored for its capability to
automatically identify and flag data corrupted by RFI.


\section{Summary}
\label{summary}

We have presented \pipe,{} a CASA-based data reduction pipeline for calibrating and imaging  VLBI data.
\pipe{} performs phase calibration using a global Schwab-Cotton fringe fit algorithm.  
A robust assessment of the S/N for each scan 
enables setting optimal fringe fit solution intervals 
for different antennas to determine post-correlation phase, delay, and rate corrections.
The amplitude calibration is based on standard telescope metadata, and a robust algorithm can solve for atmospheric opacity attenuation in the high-frequency regime. 
\pipe{} is agnostic about the details of VLBI data and can run without any user interaction, using self-tuning parameters. Its flexibility allows the user to calibrate VLBI data from different arrays, including high-frequency and low-sensitivity arrays. 
Standard CASA tasks are used for imaging and self-calibration, and
fast computing times are achieved by MPI-based CPU scalability.

In order to illustrate
the calibration and imaging capabilities of \pipe{},
we selected a 7\,mm VLBA observation of the central radio source in the M87 galaxy as a representative VLBI experiment.
We successfully applied the full end-to-end pipeline to this dataset using default parameter settings and
produced science-ready results. 
A qualitative comparison with results obtained from standard techniques with the classic VLBI software suites, AIPS and Difmap, 
has shown excellent agreement with \pipe{}.

This pipeline was used as one of the three independent
data reduction paths for 1.3\,mm EHT measurements taken during the April
2017 observations.
Calibrated data from \pipe{}, an AIPS-based reduction,
and the EHT-HOPS pipeline \citep{Blackburn2019} show a
high degree of consistency \citep{eht-paperIII}.
The calibrated measurements constitute the first scientific
data release of the EHT,
which was used to make the first image of a black hole shadow \citep{eht-paperI, eht-paperIV,eht-paperV,eht-paperVI}.\footnote{EHT data releases are available at \url{https://eventhorizontelescope.org/for-astronomers/data}.}

\begin{acknowledgements}

The authors thank Daan van Rossum for providing a flexible high-performance computing infrastructure
at Radboud University dedicated to CASA and \pipe{} testing and development.
Furthermore, we wish to thank Jose L. G\'omez for helpful comments and discussions about using \textit{tclean}
with VLBI data. The optimal set of auto-masking parameters that he determined are now used as default setting
for the \pipe{} imager.
We thank Harro Verkouter for adding new features to the jplotter program, which are improving
the data visualization capabilities of the pipeline.
This work benefited from extensive data consistency checks and useful discussions within the
Event Horizon Telescope consortium. We also thank Geoffrey C. Bower for helpful comments.
The \pipe{} source code
makes extensive use of the NumPy \citep{numpy} and SciPy \citep{scipy} python packages.

This work is supported by the ERC Synergy Grant “BlackHoleCam: Imaging the Event Horizon of Black Holes” (Grant 610058). We thank the National Science Foundation (AST-1440254). This work was supported in part by the Black Hole Initiative at Harvard University, which is supported by a grant from the John Templeton Foundation.

\end{acknowledgements}

\bibliographystyle{aa} 
\bibliography{CASApipeline1.bib}

\begin{appendix}

\section{Calibration of 7mm M87 BW0106 VLBA data}
\label{m87calib}

\begin{figure}[h]
   \centering
   \includegraphics[width=0.45\textwidth]{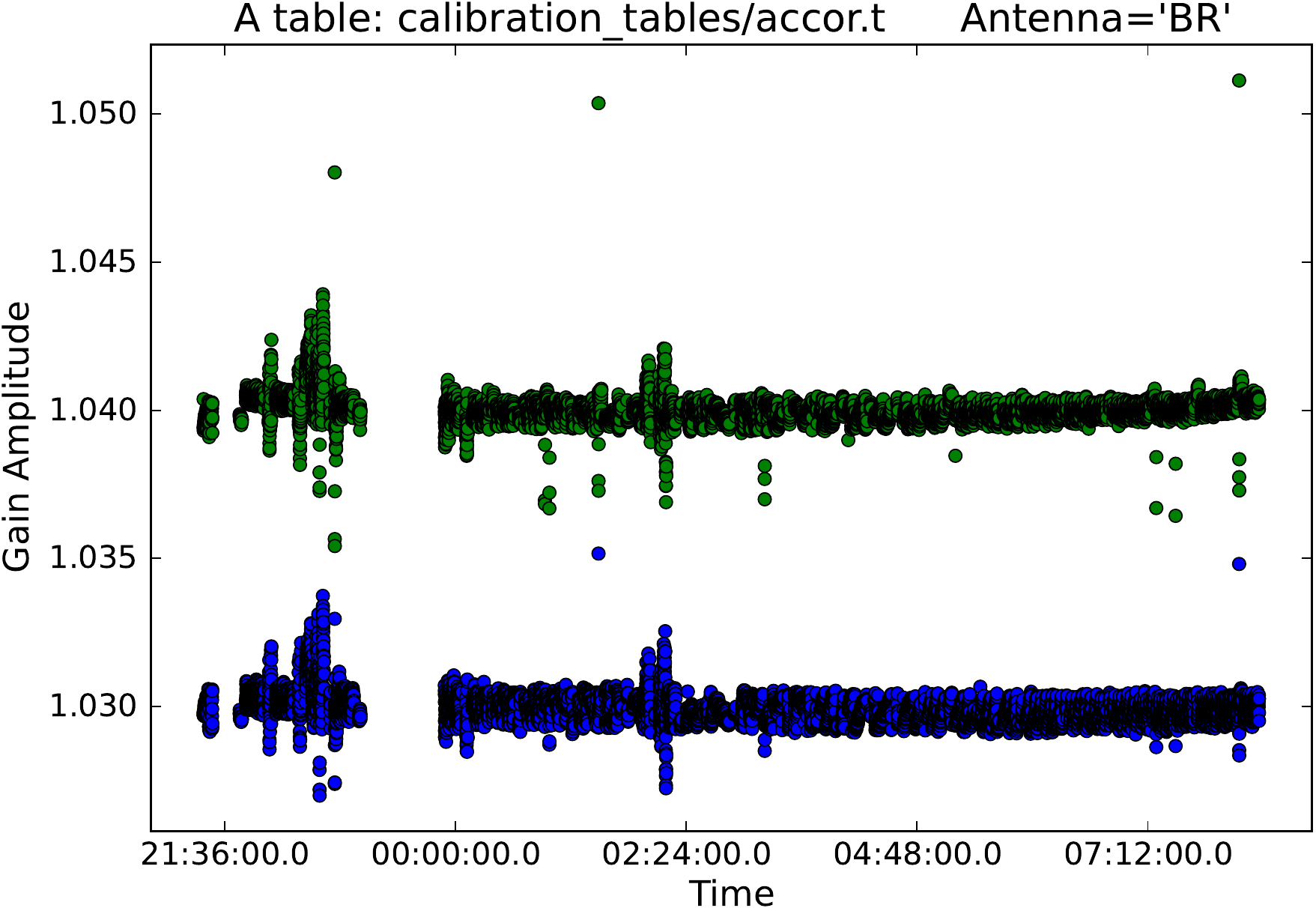}
      \caption{\textit{accor} calibration table for the LCP receiver of the Brewster antenna.
      The two groups of points correspond to two separate spectral windows, shown in different colors.
      These solutions correct for sampler threshold offsets per data integration time.
              }
         \label{vlbaaccor}
   \end{figure}

\begin{figure}[h]
   \centering
   \includegraphics[width=0.45\textwidth]{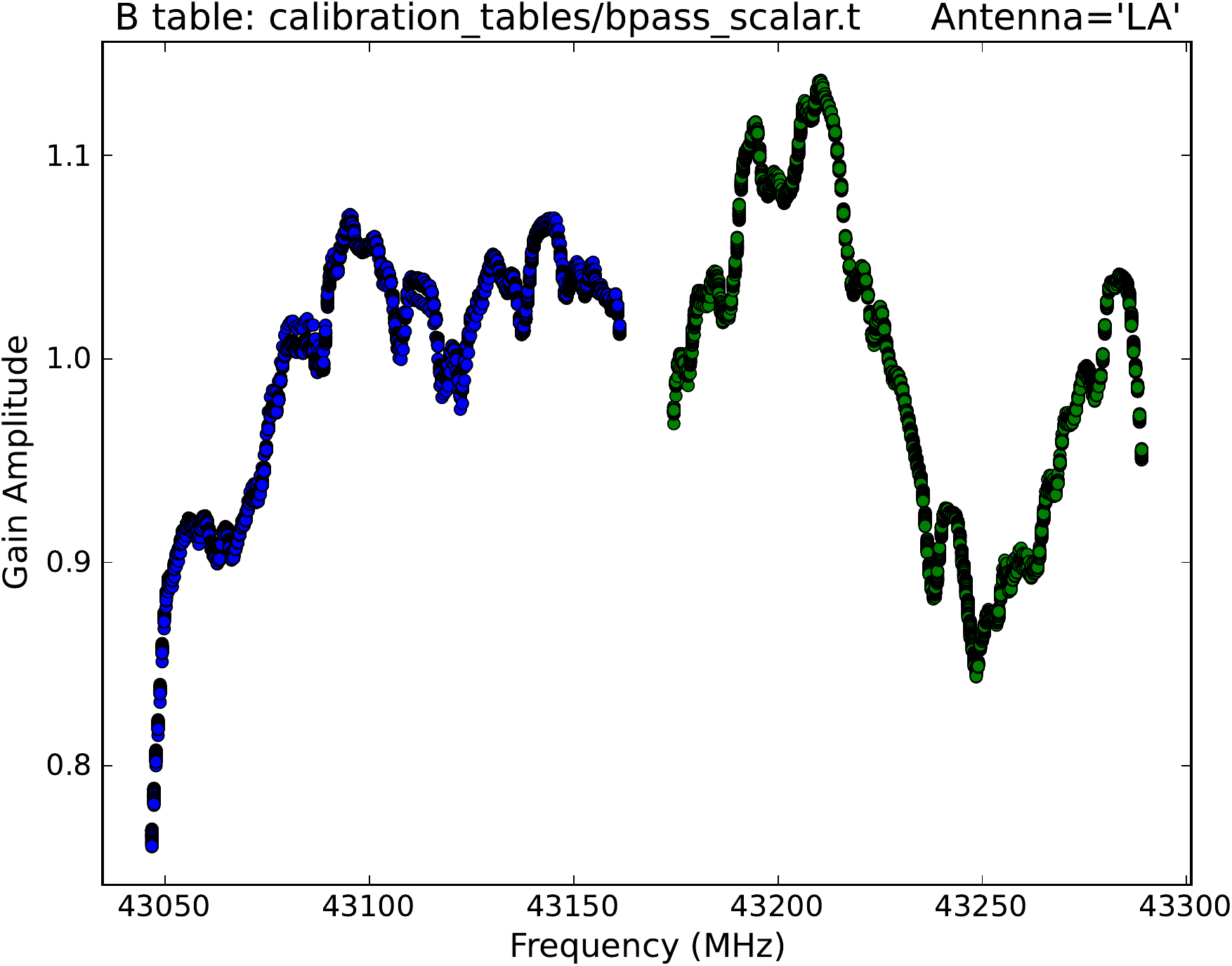}
      \caption{Scalar bandpass calibration table for the RCP receiver of the Los Alamos antenna.
               Solutions from each scan are overplotted.
              }
         \label{vlbascalarbpass}
   \end{figure}

\begin{figure}[h]
   \centering
   \includegraphics[width=0.45\textwidth]{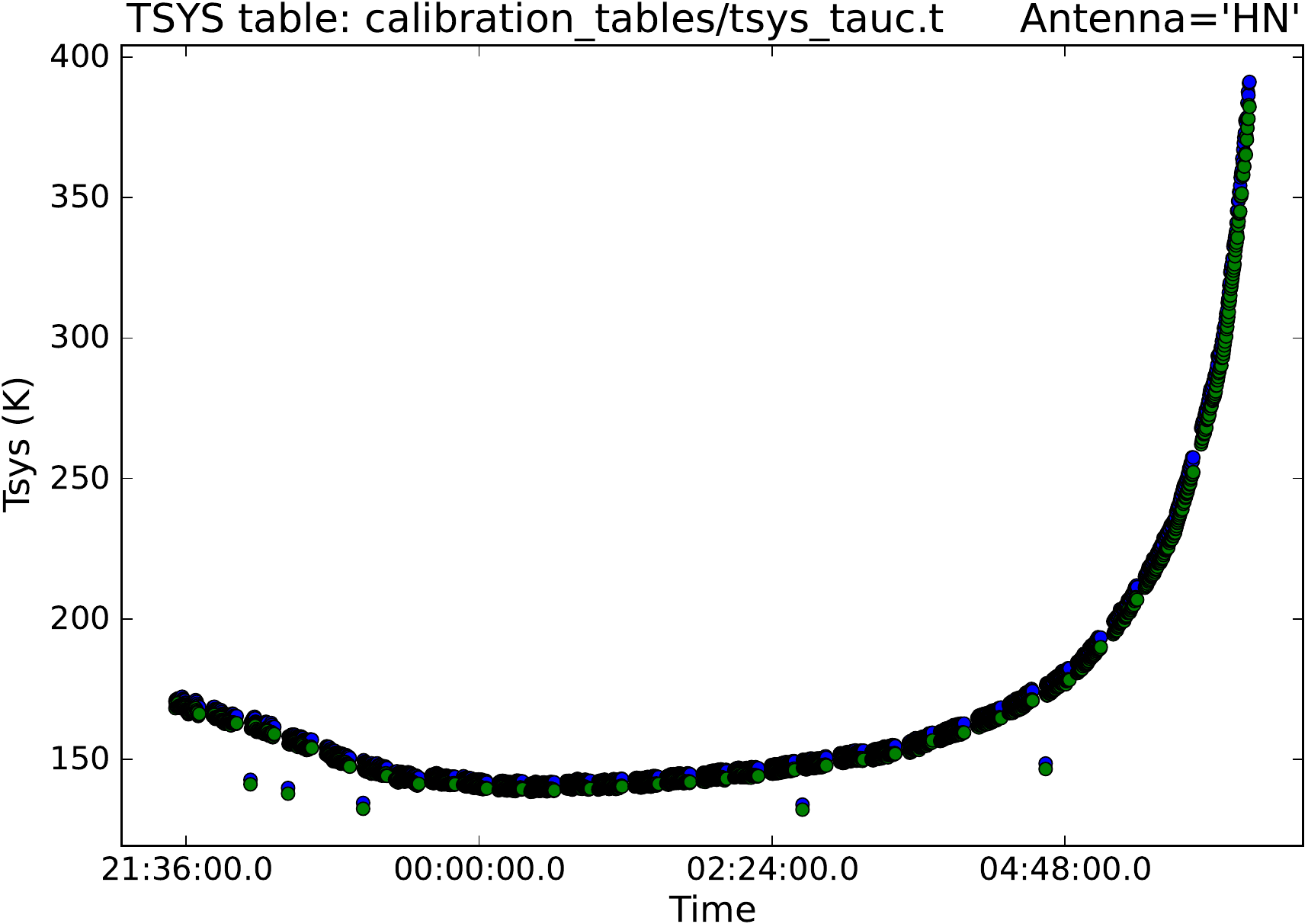}
      \caption{Opacity-corrected system temperatures for the LCP receiver of the Hancock antenna as
      M87 is tracked. This calibration table was made by \textit{gencal} using the SYSCAL data from the MS.
              }
         \label{vlbatsys}
   \end{figure}

\begin{figure}[h]
   \centering
   \includegraphics[width=0.45\textwidth]{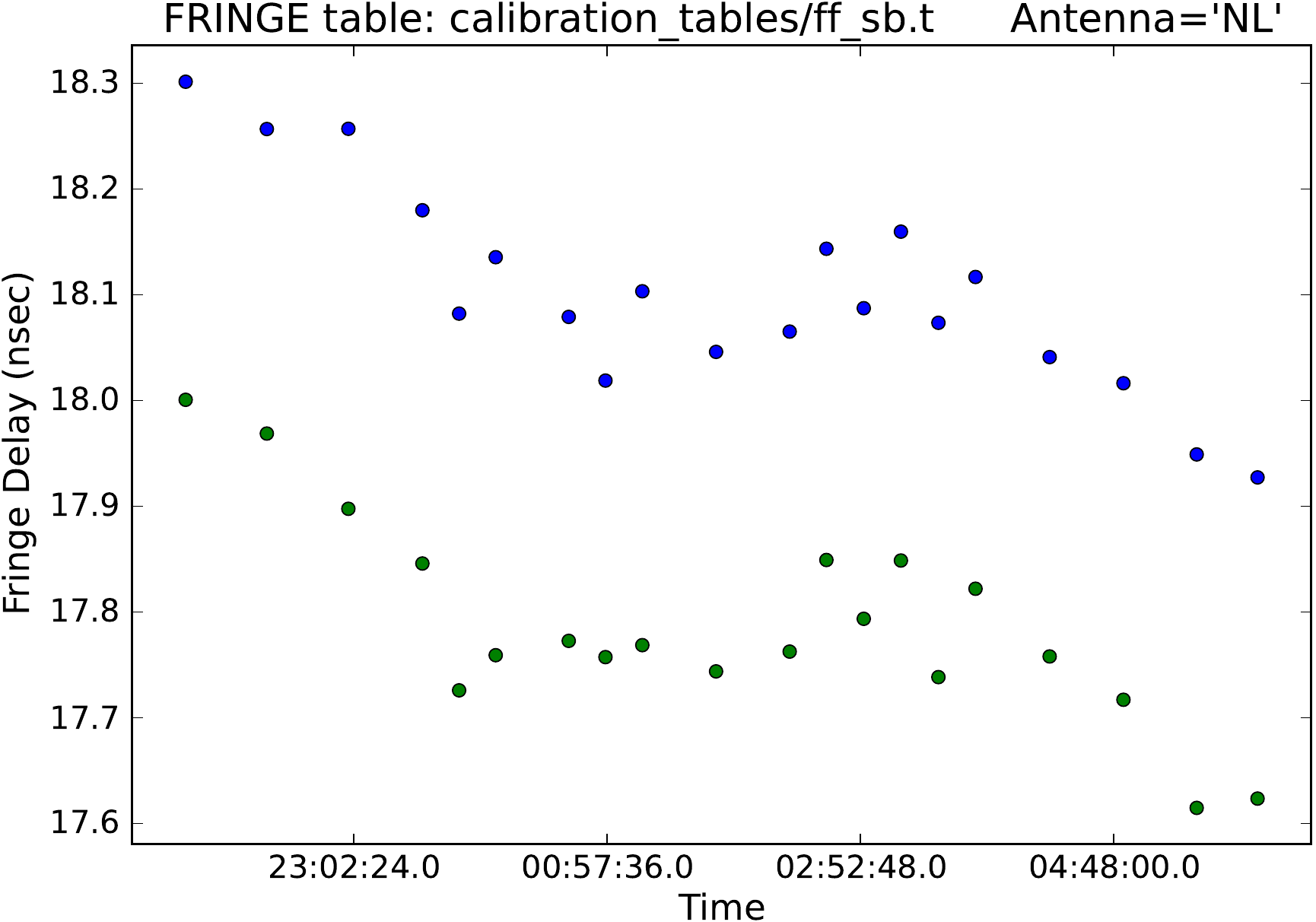}
      \caption{\textit{fringefit} calibration table solving for single-band delay offsets
      for the RCP receiver of the North Liberty station.
      The two groups of points correspond to two separate spectral windows, shown in different colors.
              }
         \label{vlbasb}
   \end{figure}

\begin{figure*}[h]
   \centering
   \includegraphics[height=0.22\textwidth]{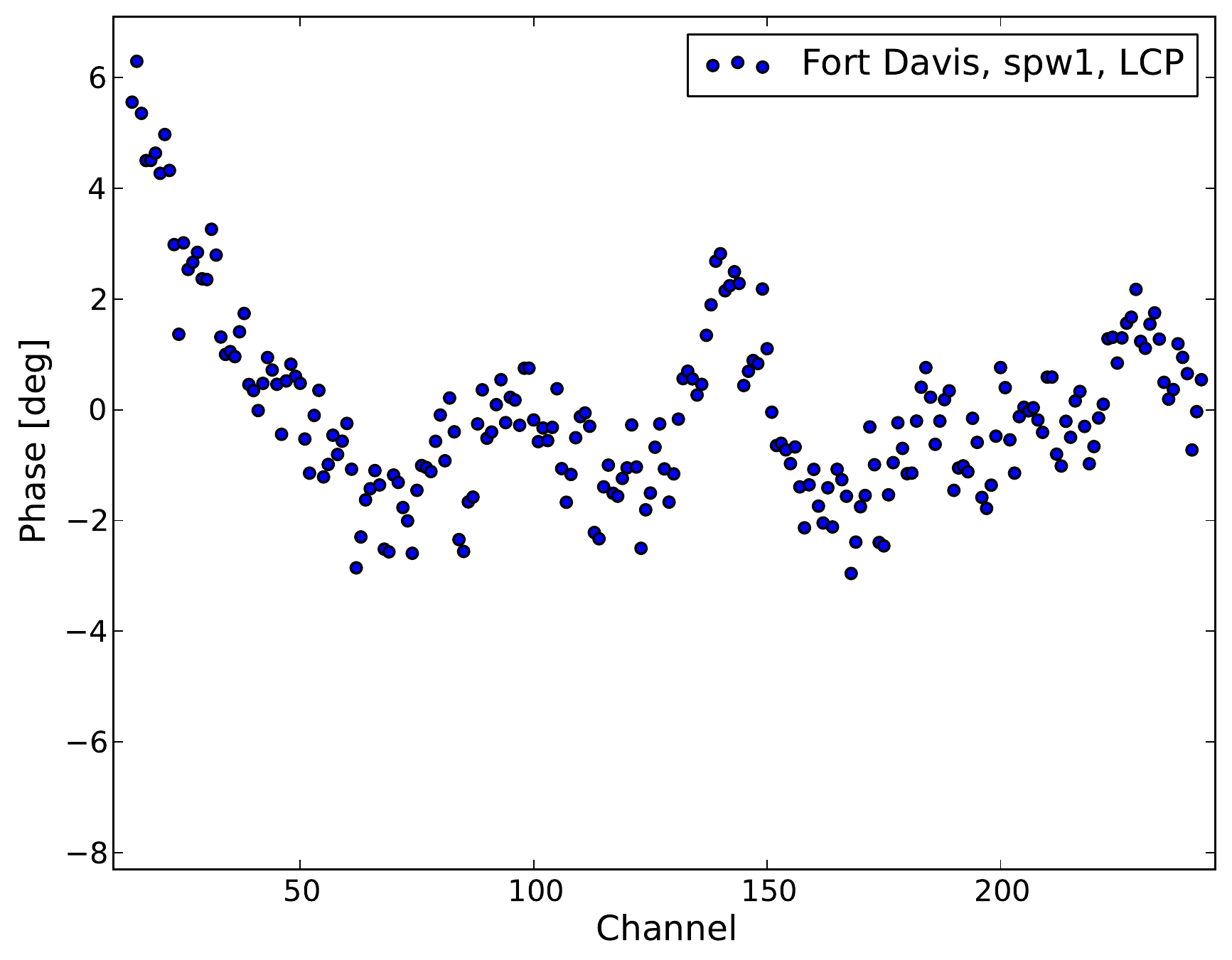} \hspace{-0.25cm}
   \includegraphics[height=0.22\textwidth]{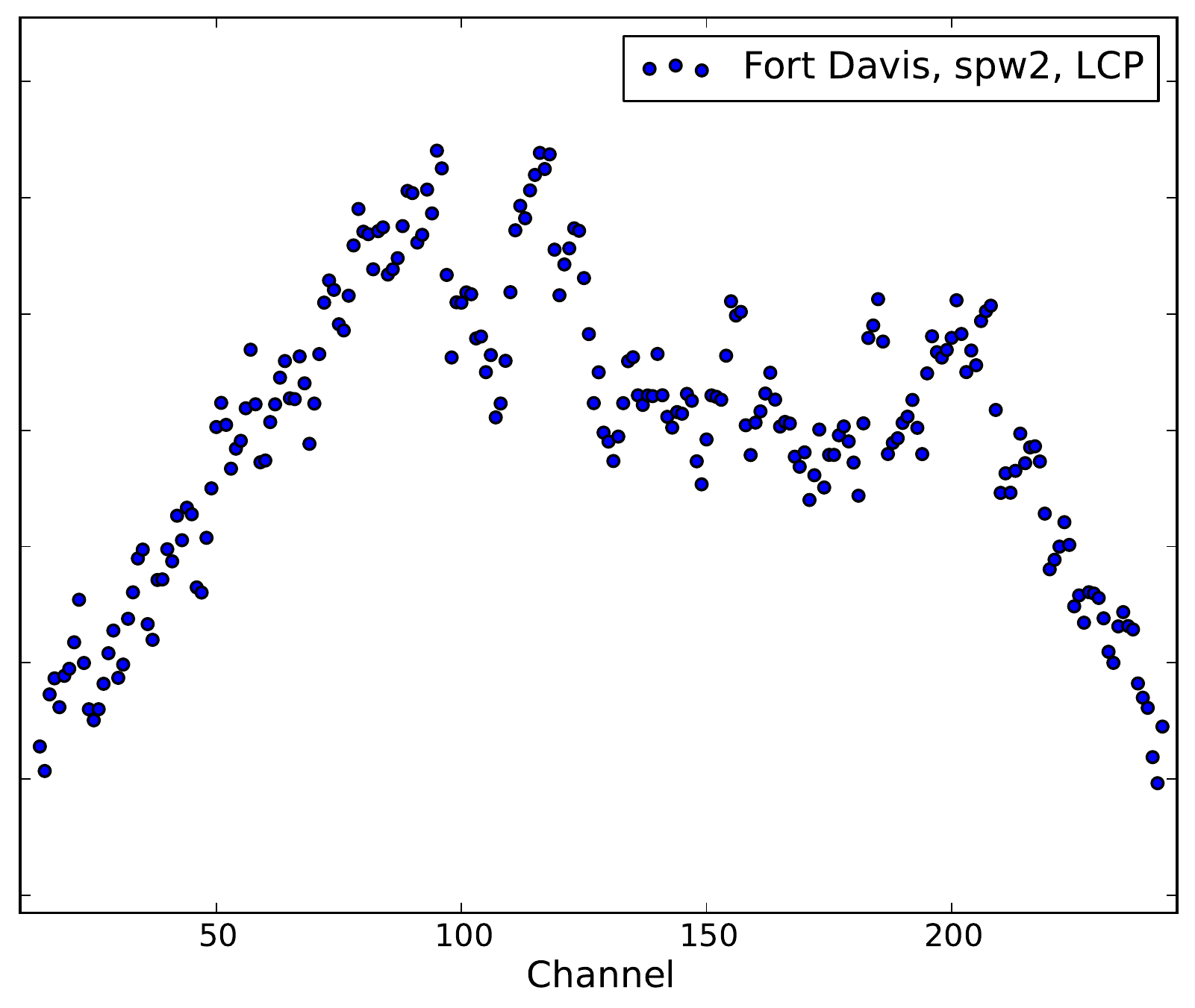} \hspace{0.7cm}
   \includegraphics[height=0.205\textwidth, trim=0 0 0 0.8cm]{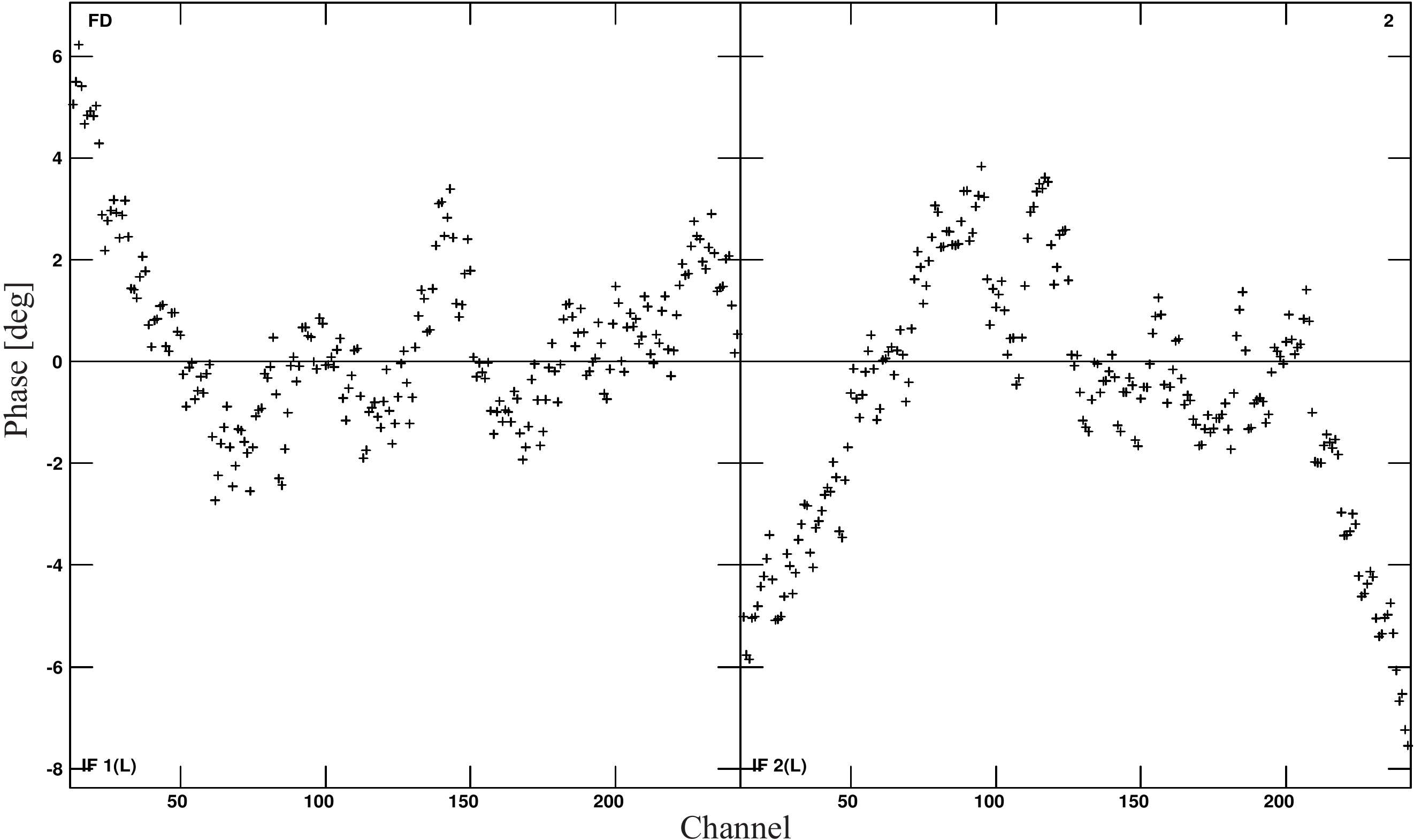}
      \caption{Complex \textit{bandpass} calibration tables solving for the phase response within
      the two spectral windows (IFs) of the Fort Davis LCP receiver as a function of frequency channel, shown
      for \pipe{} calibrated data (left panel) and 
      for AIPS calibrated data (right panel).
              }
         \label{vlbacomplexbpass}
   \end{figure*}

\begin{figure*}[h]
   \centering
   \includegraphics[height=0.3\textwidth]{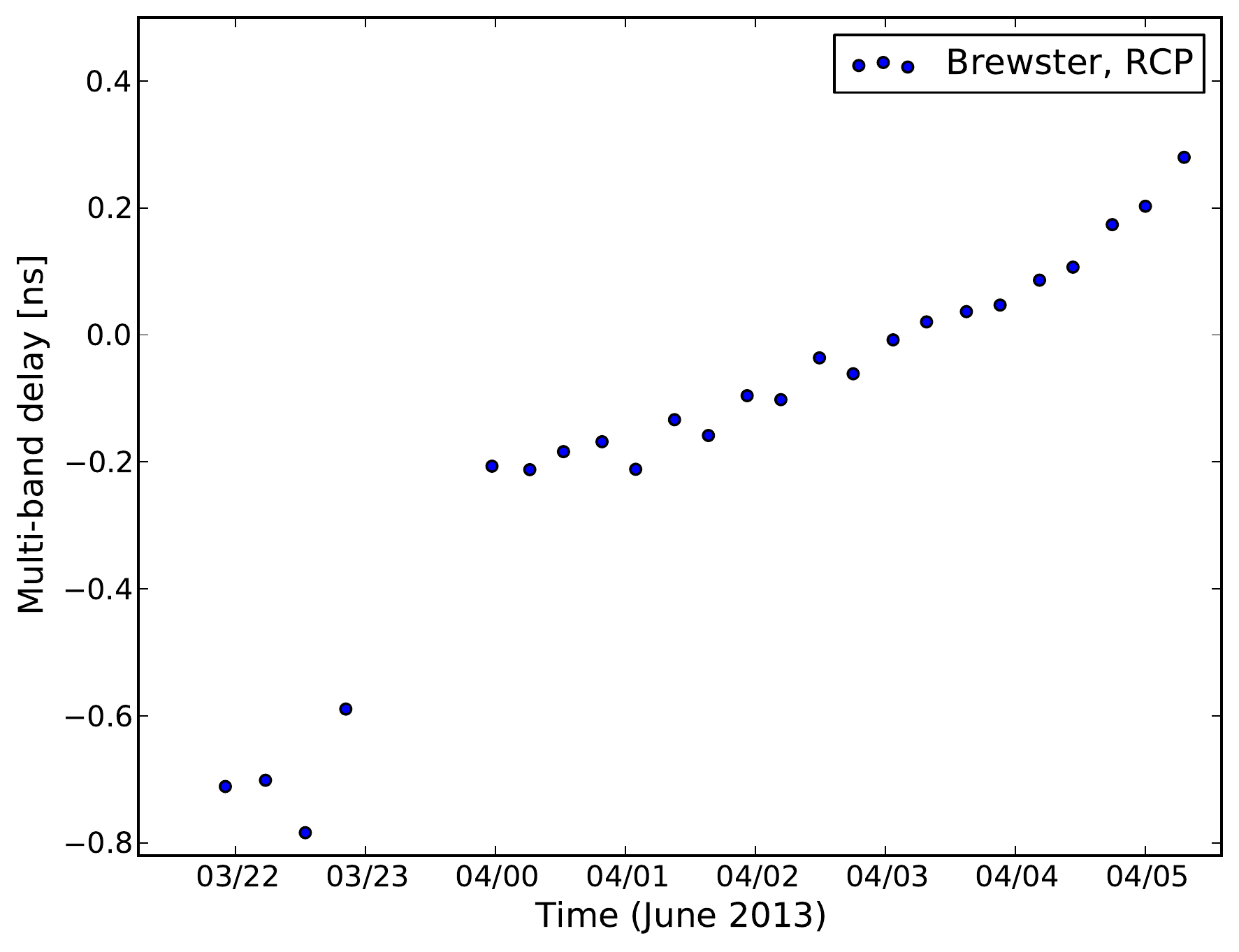} \hspace{0.7cm}
   \includegraphics[height=0.305\textwidth]{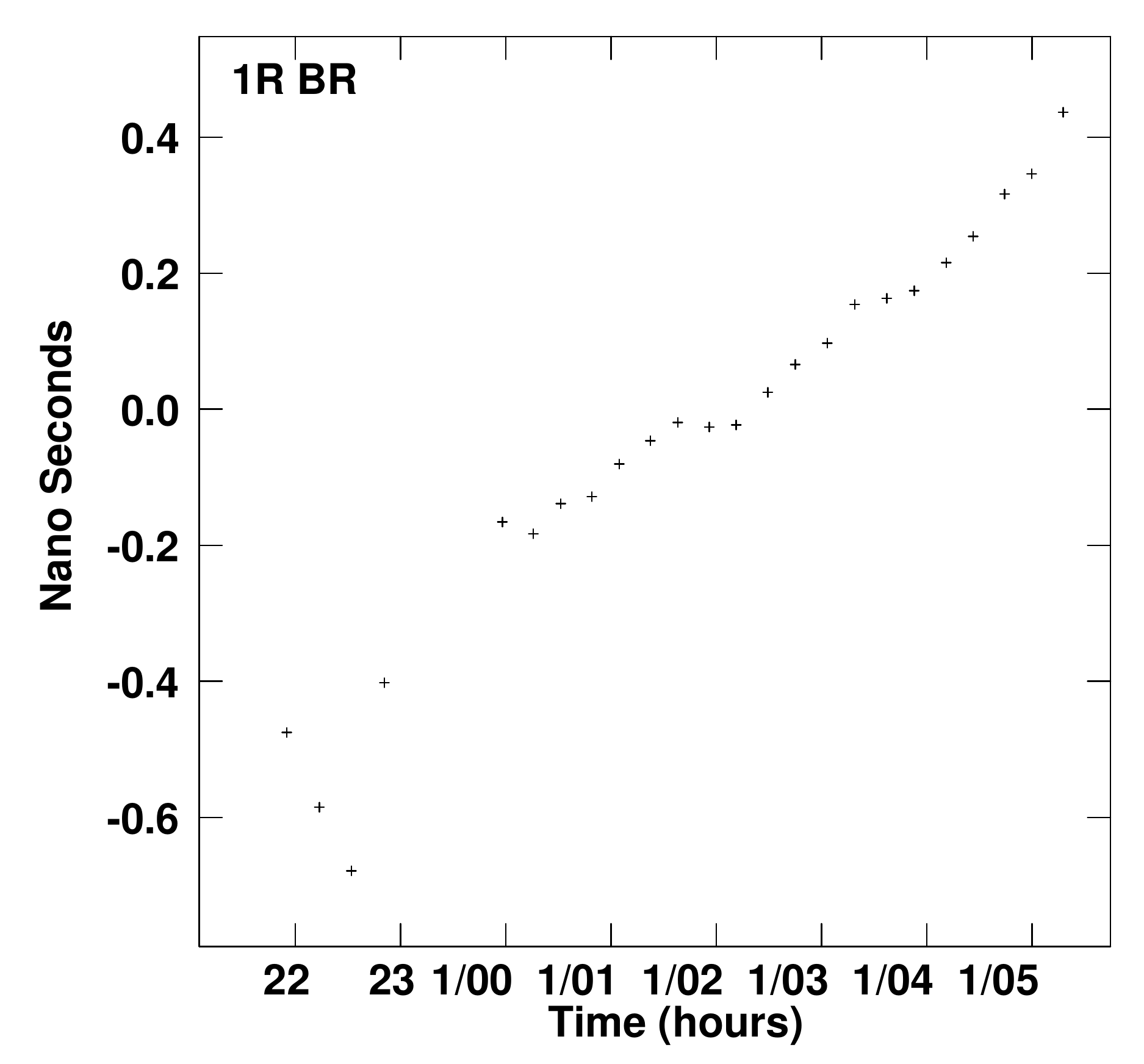}
      \caption{Scan-based multi-band fringe fit delay solutions on M87 of the RCP
      receiver of the Brewster station shown for \pipe{}
      (left panel) and AIPS (right panel).
              }
         \label{vlba-aipscasa-br}
   \end{figure*}

\begin{figure*}[h]
   \centering
   \includegraphics[height=0.3\textwidth]{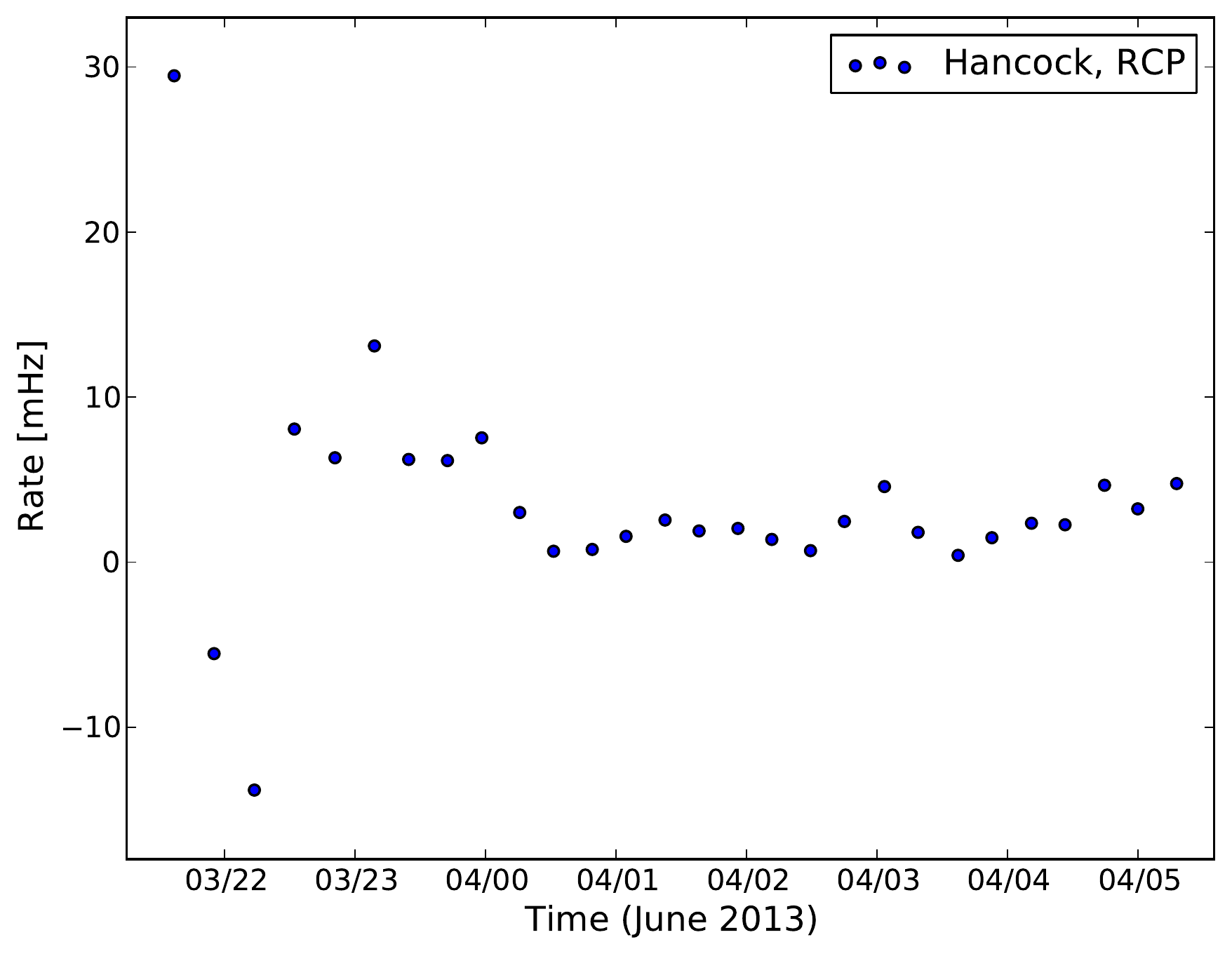} \hspace{0.7cm}
   \includegraphics[height=0.305\textwidth]{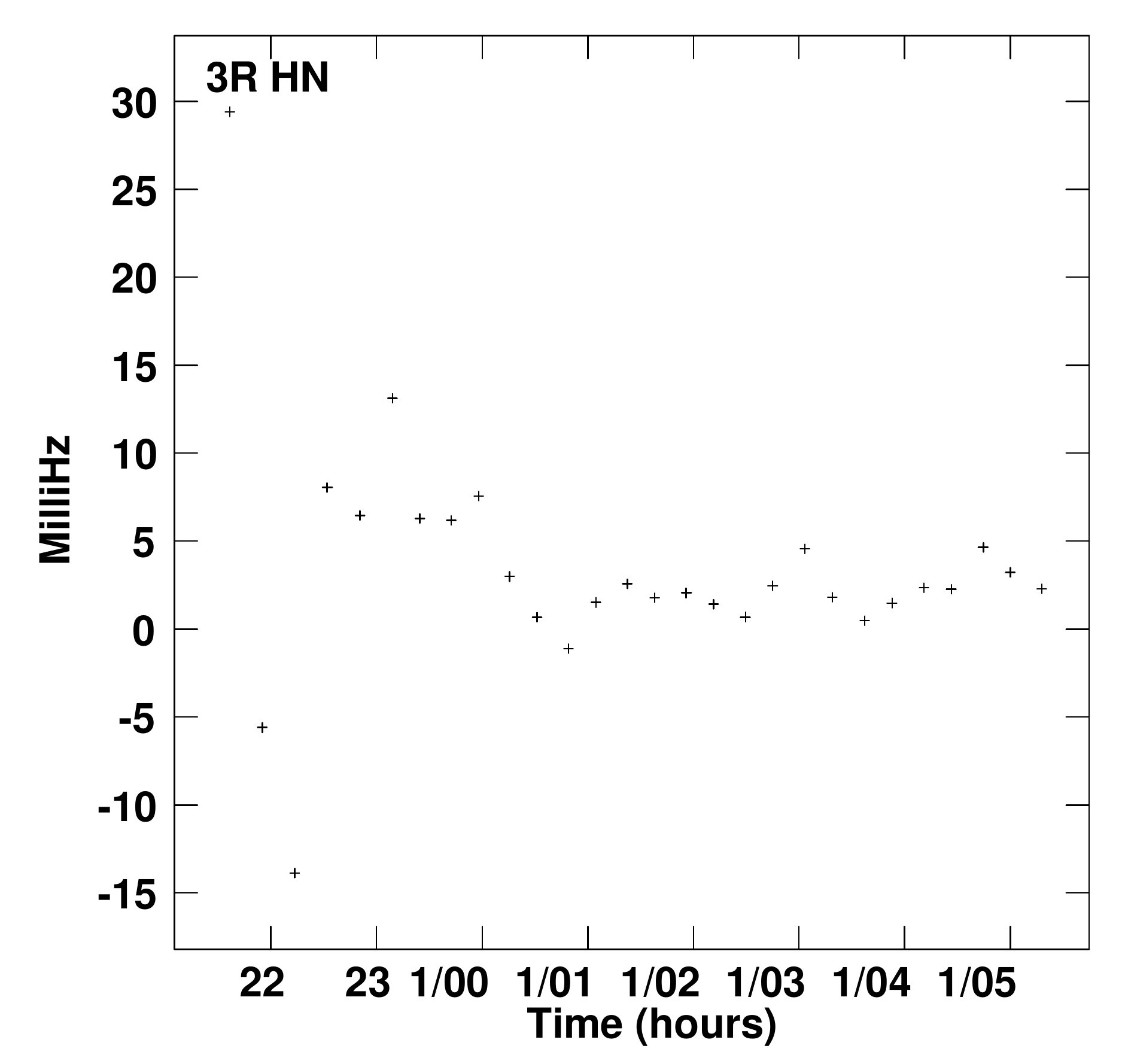}
      \caption{Scan-based multi-band fringe fit rate solutions on M87 of the RCP
      receiver of the Hancock station shown for \pipe{} (left panel) and AIPS (right panel).
              }
         \label{vlba-aipscasa-hn}
   \end{figure*}

Here, we show details of the \pipe{} and AIPS calibration of the 7mm VLBA data described in Section~\ref{verification}.
Figures \ref{vlbaaccor} to \ref{vlba-aipscasa-hn} were generated with the CASA \textit{plotcal} task (when showing \pipe{} data) as part of the pipeline default diagnostics.
All plots of calibration solutions shown here are taken from single representative antennas,
and except for the multi-band fringe fit tables presented in Fig.~\ref{vlba-aipscasa-br} and Fig.~\ref{vlba-aipscasa-hn}, they show the two spectral windows present in the data color-coded blue and green for \pipe{} data.

A calibration for sampler thresholds (Section~\ref{digicorr})
is shown in Fig.~\ref{vlbaaccor}. The corrections are very small and
extremely stable in time.
A $T_\mathrm{sys}^*$ calibration table (Section~\ref{apcal_section})
is shown in Fig.~\ref{vlbatsys} for M87. The imprint of the opacity is clearly seen as the source is tracked over a 
wide range of elevations.
Scalar bandpass solutions (Section~\ref{scalar-bpass}) are presented in
Fig.~\ref{vlbascalarbpass} for each
scan separately, showing the stability of the bandpass in time.
The stability of the instrumental phase and delay calibration (Section~\ref{instrphdela}) is
illustrated by the constant delay offsets between the spectral windows shown in Fig.~\ref{vlbasb}.
Per-channel phase bandpass solutions (Section~\ref{bpass}) are plotted in
Fig~\ref{vlbacomplexbpass} for both \pipe{} and AIPS data. The combined solutions
from 3C279 and OJ287 have enough S/N to solve for distinct bandpass responses of a few degrees
in a good agreement between AIPS and CASA.
Multi-band fringe fit solutions over full scans (Section~\ref{mbphsci})
are shown for delays in Fig.~\ref{vlba-aipscasa-br} and rates in Fig.~\ref{vlba-aipscasa-hn}
for both \pipe{} and AIPS. The same fringe solutions are recovered by the two calibration packages.

\newpage
\clearpage

\section{Calibration weights}
\label{incrcalib_and_wt}
The \pipe{} calibration framework introduced in Section~\ref{picalib} performs incremental calibration steps based on the CASA calibration scheme.
The default choice is to solve for the amplitude calibration solutions first, as these will adjust the WEIGHT column.
As follows, the weights of the data within each solution interval will be adjusted
when the amplitude solutions are applied on-the-fly for the phase calibration steps.
The adjustment of the weights is based on the quality of the data as indicated
by the magnitudes of the applied amplitude gains: data that need large correction
factors are considered worse and will be down weighted. 
In the time domain, weights are primarily adjusted by
system temperatures.
For phased arrays, short-timescale phasing efficiencies should enter into the
$T_\mathrm{sys}$ estimates. For single dishes, total power measurements can be used
to modulate the system temperature by atmospheric variations within scans.
Variations of gain curves have a negligible effect on the weights within the solution intervals.
Along the frequency axis, the weights are affected by the amplitude bandpass solution (Section~\ref{scalar-bpass}).

\section{Antenna mount types}
\label{mounttypes}
For the most common telescope designs used in VLBI observations, the rotation of the receiver polarization feeds with respect to
the source causes a phase evolution $\chi$ as a function of time.
Before fringe fitting, this simple geometric effect should be corrected, which is done on-the-fly within CASA. This is particularly important
for a proper polarization calibration (Section~\ref{polcalsect}).
The feed rotation can be attributed to phase rotations caused by telescope focus positions ($\chi_f$) and mount configurations ($\chi_m$).
Cassegrain, Prime, and Gregorian foci do not cause time-dependent phase rotations:
\begin{equation}
\label{eq-foci1}
\chi_f^\mathrm{Cassegrain} = \chi_f^\mathrm{Prime} = \chi_f^\mathrm{Gregorian} = 0 \; .
\end{equation}
For Nasmyth focus positions, a phase rotation occurs due to the changing elevation angle $E$ of a source that is tracked during an observation:
\begin{equation}
\label{eq-foci2}
\chi_f^\mathrm{Nasmyth} = \chi_f^\mathrm{Nasmyth}(E) = \pm E \; ,
\end{equation}
where the plus sign indicates the evolution for Nasmyth-right mounts and the minus sign indicates the evolution for Nasmyth-left mounts.
Regarding telescope mount (or drive) types, the most common configuration is an altitude-azimuth system, which causes a phase rotation
due to the parallactic angle $\chi_p$ as \citep{Cotton1993}:
\begin{equation}
\label{parang}
\chi_m^\mathrm{Alt-Az} = \chi_p \equiv \tan^{-1} \left( \frac{\cos(\varphi) \sin(h)}{\sin(\varphi) \cos(\delta) - \cos(\varphi) \sin(\delta) \cos(h)} \right)  \; ,
\end{equation}
with $\varphi$ the station latitude, $\delta$ the declination of the source, and $h$ the source hour angle.
For east-west mounts, a rotation with the coparallactic angle $\chi_c$ occurs as
\begin{equation}
\label{coparang}
\chi_m^\mathrm{East-West} = \chi_c \equiv \tan^{-1} \left( \frac{\cos(h)}{\sin(\delta) \sin(h) }\right)  \; .
\end{equation}
For equatorial mounts, no phase rotation occurs:
\begin{equation}
\label{equatorial}
\chi_m^\mathrm{Equatorial} = 0 \; .
\end{equation}
The total feed rotation angle is then
\begin{equation}
\label{totalfeedrotation}
\chi^\mathrm{\textit{focus, mount}} = \chi_f^\mathrm{\textit{focus}} + \chi_m^\mathrm{\textit{mount}} \; .
\end{equation}

\section{Fringe fitting in \pipe{}}
\label{ffbasics}

   \begin{figure}[ht]
   \centering
   \includegraphics[width=0.3\textwidth]{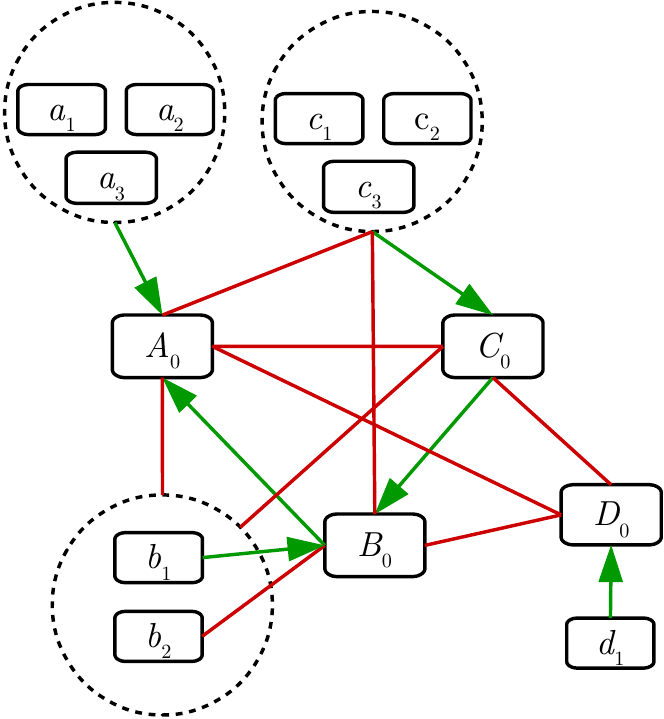}
      \caption{Schematic representation of fringe-detections across a VLBI array to exemplify the
               exhaustive fringe-search algorithm implemented in \pipe{}. $A_0, B_0, C_0, \text{and } D_0$
               represent prioritized reference stations, and all $a_i, b_i, c_i,\text{and } d_i$ are other stations
               in the array. Green arrows indicate successful fringe-detections (with high enough S/N) to a specific
               reference station, and red lines indicate non-detections. Connecting lines to dashed circles
               around groups of stations represent fringes to each individual station within the circle.
               Not all possible lines are drawn; all omitted connections are considered nondetections.
               The intra-scan re-referencing of the exhaustive fringe-search is accomplished through the connection
               of fringe detections between the reference stations.
              }
         \label{fig:exhaustive-fringe-serach}
   \end{figure}

Fringe fitting is used as the primary phase-calibration task by \pipe{}
(Section~\ref{phasecal}).
The basic principle is to first use a Fourier transform on the visibilities as a function of time into the rate space
and as a function of frequency into the delay space using an FFT \citep{whitebook}.
If the observed source has been detected
on a specific baseline, strong signals will be present at the post-correlation rate and delay offsets.
The strengths of these signals are characterized by their S/N as given by their amplitude $A_\mathrm{fringe}$ in the Fourier domain
over the noise $\sigma$: $\mathrm{S/N}_\mathrm{fringe} = A_\mathrm{fringe} / \sigma$.
The signal search is made over discrete rate and delay grid cells. Within specified search windows, the cell with the maximum
amplitude is chosen.
Under the assumption that the signal is present in only one
of the cells, a false detection occurs if the amplitude in a cell without signal exceeds the signal amplitude
\citep[see][for details]{bluebook}. Following \citet{Rogers1995}, the probability of false detection $P_\mathrm{error}$ can be estimated as
\begin{equation}
P_\mathrm{error} \approx n e^{ - \mathrm{S/N}_\mathrm{fringe}^2/2} \; ,
\end{equation}
where $n$ denotes the number of discrete rate and delay cells within the search windows.
It follows that the FFT $\mathrm{S/N}_\mathrm{fringe}$ serves as a good indicator to distinguish detections from nondetections
and that a priori knowledge about the approximate fringe location in the delay-rate search space is important in the low S/N regime.

In practice, the FFTs are obtained on all baselines to a specified reference station.
Reference antennas should be sensitive and central in the array to maximize the S/N on each 
connecting baseline.
An S/N cut $\mathrm{S/N}_\mathrm{min}$ is used for $\mathrm{S/N}_\mathrm{fringe}$. This determines whether an acceptable solution or detection can be obtained for
a baseline between a station $s$ and reference antenna $R$. If $\mathrm{S/N}_{\mathrm{fringe}}(s,R) < \mathrm{S/N}_\mathrm{min}$, a flagged
solution will be written for $s$.
In global VLBI experiments, a single reference $R$ often does not suffice to obtain fringes to every other station in the array.\footnote{
Typically, the S/N is higher on shorter baselines because the source flux is measured on a larger scale.}
To overcome this problem, \pipe{} employs an exhaustive fringe search, where fringes to a
prioritized list of reference stations $R_0, R_1, R_2, \text{etc.}$ are searched.
A station $s$ can be calibrated when fringes on a baseline $s$\,--\,$R_{i\neq0}$ are found, even
if the signal between $s$ and the primary reference station $R_0$ is too weak, as long as
fringe solutions from $R_{i\neq0}$ can be re-referenced to $R_0$.
For each scan, the first available station from the prioritized list is selected as a primary 
reference antenna. Fig.~\ref{fig:exhaustive-fringe-serach} illustrates the
exhaustive fringe search principle. In the example shown, direct fringes to the primary
reference $A_0$ are found only for the stations $a_i$. Antenna $b_1$ can only be calibrated
through fringes found to $B_0$, re-referenced to $A_0$, with the fringe relation between $A_0$
and $B_0$. $b_1$ cannot be calibrated because no fringes to any reference station are found.
The $c_i$ stations can only be calibrated through $C_0$ to $B_0$ to $A_0$ because no fringes
are found in the more direct $C_0$\,--\,$A_0$ connection. $D_0$ and $d_1$ cannot be calibrated
even if fringes are found on their connecting baseline because they cannot be connected to
the rest of the array.
In the end, all station-based fringe solutions are referenced to a specific primary station for each scan.
The CASA \textit{rerefant} task is used to re-reference all these solutions to a single station for all scans in a
VLBI experiment to ensure phase continuity.
The default choices for reference stations are given in the online documentation.
Re-referencing across scans can lead to small phase errors because the phase of the 
primary reference station cannot be tracked during the time it could not observe.
For polarization experiments, a single primary reference station can be enforced for each scan
to overcome this problem, while still allowing for exhaustive fringe searches within scans.

All FFT solutions with sufficient S/N are passed as initial estimates to a least-squares solver
to obtained refined station-based calibration results with a subgrid resolution \citep{Schwab1983}.
The least-squares procedure will tune the station gains such that the data match an assumed source model.
In most cases, the source model is unknown at the stage of data calibration, so that a point source model is assumed.
Station gains are therefore adjusted such that baseline phases to the reference station are steered toward zero.
The true source structure is still encoded in the closure phases \citep{Jennison1958},
which are unaffected by the station-based calibration unless the data are averaged over long time periods.
Fringe solutions can be obtained for each receiver polarization channel over a set
integration time and over a set frequency range.

\section{Phase-referencing in \pipe{}}
\label{phrefastrom}
For cm observations, the modest atmospheric influence warrants the use of phase-referencing.
With this technique, fringe solutions from a nearby calibrator are transferred to the science target.
This enables the calibration of weak science targets, which would be too weak to be
detected on their own.
Phase-referencing is one of the possible phase calibration paths implemented
in \pipe{} (Section~\ref{mbphsci}).
Typically, this is done by interleaving short calibrator scans with science target scans. The length
of these scans should not exceed the atmospheric coherence time.
Before the fringe transfer, the solutions are first filtered to
remove flagged (failed) detections, and smoothed to 
a single value per scan, antenna, and spectral window.

For astrometry experiments, phase-referencing is essential to obtain accurate positions of the science
targets relative to the calibrator sources. When the source position is of no interest and when the science source is strong enough, it
can be fringe fit to correct for residual atmospheric
effects after the phase transfer. 
For high-frequency VLBI experiments, atmospheric contributions to the phases prohibit phase-referencing
unless in-beam calibrators are available.

\section{MPI scalability test case}
\label{mpibench}
A quick single-core versus multi-core
comparison of an eight-hour-long 256\,MHz bandwidth VLBA experiment
(Section~\ref{verification}) demonstrated the significance
of the CPU scalability: It took \pipe{} two hours and three minutes
to process the entire dataset sequentially, which was reduced to
34 minutes when 14 CPU cores were used in parallel. For the parallelized
fringe fitting steps, a speed-up factor of 6.7 was achieved.
This test was made with a Intel(R) Xeon(R) CPU E5-2690 v4 at 2.60\,GHz
and a Samsung 960 Pro SSD hard drive connected to four gen-3 PCIe lanes.

However, a general MPI benchmark test has a limited significance because 1) not all
CASA tasks are parallelized, 2) computing speeds are heavily affected by disk
I/O (because the datasets are too large to be stored in memory),
and 3) large differences in data quality across scans can significantly
slow down the multi-CPU core speed-up (because a few bad scans can dominate the total 
computing time).

\end{appendix}

\end{document}